%% file: afroz.tex
\documentclass[reprint, amsmath, amssymb, aps,pre,twocolumn,10pt]{revtex4-2}

\usepackage[normalem]{ulem}
\usepackage{graphicx}
\usepackage{dcolumn}
\usepackage{bm}
\newcommand{\fw}{3.2}

\begin{document}

\title{Analysis of Granular Flow in a Wedge-Shaped Hopper}

\author{Afroz Momin}
\author{Devang Khakhar}
\affiliation{Department of Chemical Engineering, Indian Institute of Technology Bombay, Powai, Mumbai 400076, India}

\date{\today}

\begin{abstract}
We study the velocity and stress distribution for a planar granular flow in a wedge-shaped hopper using Discrete Element Method simulations. Periodic boundary conditions are used in the direction normal to the plane of the flow to simulate a layer system. A parametric study is carried out varying system parameters (orifice width, wedge angle, wall friction and hopper height) and particle properties (particle diameter and particle friction). Scaling relations are obtained for the velocity and stress ratios. The data indicate that the stresses are not dependent on the shear rate, as in the Mohr-Coulomb rheology, however, the friction coefficient varies spatially, and the coaxiality condition is violated in some regions of the flow. A theory based on the scaling relations is presented, which gives good predictions for all the cases studied. Based on the theory, an expression for the mass flow rate, incorporating the effect of system parameters and particle properties is derived, which gives a close match with the computed mass flow rates for all the cases.
\end{abstract}

\maketitle

\section{Introduction}
Hoppers are simple, but very useful elements of most granular processing operations in industry \cite{nedderman1992statics}. The stress distribution and the flow rate from hoppers is of considerable practical importance, and has been the subject of much prior research starting with the works of Hagen \cite{tighe2007pressure} and \citet{janssen1895versuche}. Hagen \cite{tighe2007pressure} showed that the mass flow rate of grains from a cylindrical bin with a circular opening at its base is given by
\begin{equation}\label{eq:hag}
\dot{m} = C(D-\delta)^{5/2}
\end{equation}
where $D$ is the orifice diameter and $\delta$ is a constant of the same order as the particle diameter. Later, \citet{beverloo1961flow} validated this equation by means of extensive experiments and found $\delta = kd$, with $k \approx 1.4$. \citet{janssen1895versuche} showed theoretically that the stress at the base of a hopper, in the absence of flow, is independent of the height of particles filled in the hopper, and attributed this to the weight of the particles being partially carried by the frictional side walls. Following these early works, a number of theoretical analyses of the flow in hoppers have been carried out, and we briefly review them below.

\citet{jenike1964} carried out detailed analyses of the flow in hoppers by experiments and theory, and several studies followed a similar approach. The primary postulates in Jenike’s theory \cite{jenike1964} are: (i) constant density, (ii) rheology determined by the Mohr-Coulomb condition \cite{jenike1959plastic} extended to flowing systems, and (iii) coaxiality of the stress and strain rate. These are given in more detail in Sec.~\ref{sec:the}. \citet{jenike1964} assumed in the analysis, that the inertial terms in the momentum balance equation are negligible, and, as a consequence, the velocity field is indeterminate, since the stresses are independent of the velocity. 

\begin{figure}
\begin{center}
\includegraphics[width=1.2in]{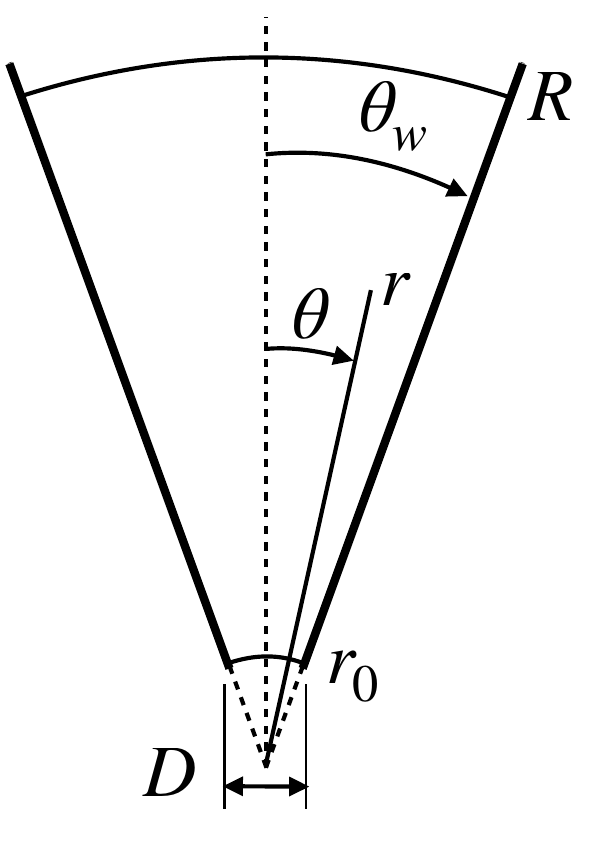}
\caption{Schematic view showing the geometry of the wedge shaped hopper, with orifice width, $D$, and wedge angle, $\theta_w$. The cylindrical coordinate system ($r,\theta$) used in the analysis is shown, along with the radial positions of the exit plane ($r_0$) and the free surface of the material ($R$).}\label{fig:sch}
\end{center}
\end{figure}

\citet{savage1965mass} applied Jenike's approach \cite{jenike1964}, but including inertial terms, to a simplified system: a wedge-shaped hopper, shown schematically in Fig.~\ref{fig:sch}, with frictionless walls and radial gravity. Inclusion of the inertial terms enables the determination of the velocity field. A closed form solution of the governing equations for the simplified system is obtained, assuming the stress to be zero at the free surface and the exit. The theory predicts a larger mass flow rate than experimental measurements. \citet{momin2025granular} carried out detailed DEM simulations of the same system, and found that the \citet{savage1965mass} theory predicts the stress distribution in the hopper very well, but the predicted flow rate is higher than that obtained from simulations. The mass flow rate is overestimated because the stress at the exit is finite and not zero, as assumed by \citet{savage1965mass}. They also found that the Mohr Coulomb rheology \cite{jenike1959plastic} is valid in most of the hopper but deviates near the exit, where it is better described by the $\mu(I)$ rheology \cite{jop2006}.

Extensions of the \citet{savage1965mass}-\citet{jenike1964} approach have been proposed to analyze the flow in wedge-shaped hoppers with frictional walls. \citet{walker1966} and \citet{walters1973} presented approximate solutions to the \citet{jenike1964} model, using cross-sectional averaging and neglecting particle inertia. \citet{horne1976} provided a solution based on the method of characteristics, again neglecting particle inertia. \citet{brennen1978} proposed a perturbation analysis of the system for small wall friction, and obtained corrections to the \citet{savage1965mass} solution. \citet{kaza1982} also presented a perturbation analysis for the problem. \citet{savage1981gravity} presented an analysis including wall friction to predict the mass flow rate, which matched experimental results. \citet{jyotsna1997} used the frictional-collision rheology to analyze the problem, and \citet{prakash1988} applied the critical state theory to consider effects of compressibility. \citet{gremaud2004} computed the stress field in different hopper geometries using the von Mises yield criterion, but neglected inertia in the stress balance equations. 

A number of previous works have focussed on the flow in hoppers, by means of experiments and DEM simulations  \cite{choi2005velocity, vivanco2012dynamical, staron2012, Janda2012, lin2015, Rubio2015, Borzsonyi2016, Aussillous2017, Bhateja2020, darias2020hopper, wang2021shape, mehdizad2021, Zou2022, gans2024discharge,yogi2026gravity}. The studies present results for velocity scaling, mass flow rate and rheology. In a recent study, the particle acceleration in the exit region is found to be larger than gravitational acceleration, and is attributed to the high normal stress gradients in the exit region \cite{yogi2026gravity}.

Despite the rigorous analyses reported in the large body of previous work, a detailed validation of the theories is lacking. Comparisons are largely limited to measurements of mass flow rate from the hopper and velocity and stress measurements at the walls of the hopper. This is because experimental measurements of stress and velocity fields in the interior of hopper flows are very difficult to make. An alternative is to use Discrete Element Method (DEM) simulations, which allow the calculation of velocity and stress fields, and show good agreement with experimental measurements. An example is the recent work \citet{yogi2026granular}, in which a good match between experimental and DEM simulation results is shown for the velocity distribution in a quasi-2D wedge shaped hopper.

In this work, we carry out DEM simulations to characterize the velocity and stress fields in a wedge-shaped hopper, shown schematically in Fig.~\ref{fig:sch}.  Simulations are carried out for different hopper parameters (orifice width, wedge angle, wall friction coefficient) and varying particle properties (coefficient of friction, diameter). The major objectives of the work are to characterise in detail the flow and stress distributions in the system, and to compare the predictions of a theory for the flow, presented below, with the DEM simulation results. The flow in the system is non-viscometric, and another objective of the work is to understand the rheology of the granular material in the flow. The general theory for the flow is presented next  (Sec.~\ref{sec:the}) followed by computational details (Sec.~\ref{sec:sim}). Results are presented in Sec.~\ref{sec:res} and conclusions of the study in Sec.~\ref{sec:con}.

\section{Theoretical formulation}\label{sec:the}
In this section, we present the the governing equations for planar granular flow in a wedge-shaped hopper with frictional side walls, shown in Fig.~\ref{fig:sch}. We first present the governing equations for the flow considering the Mohr-Coulomb rheology \cite{jenike1959plastic} and then the $\mu(I)$ rheology \cite{jop2006}. 

In the formulation given below, which closely follows \citet{jenike1964} and \citet{savage1965mass}, we assume the flow to be steady and planar in the $r$-$\theta$ plane, the bulk density, $\rho$, to be constant and the rheology to be given by the Mohr-Coulomb condition \cite{jenike1959plastic}. The continuity equation in this case is
\begin{equation}
\frac{\partial}{\partial r}(r v_r) + \frac{\partial v_\theta}{\partial \theta} = 0,
\end{equation}
and the momentum balance equations are
\begin{equation}
\begin{split}
    \rho &\left\{ v_r \frac{\partial v_r}{\partial r} + \frac{v_\theta}{r} \frac{\partial v_r}{\partial \theta} - \frac{v_\theta^2}{r} \right\}=\\ &-\frac{\partial \sigma_{rr}}{\partial r} - \frac{1}{r} \frac{\partial \sigma_{r\theta}}{\partial \theta} + \frac{\sigma_{\theta\theta} - \sigma_{rr}}{r} - \rho g \cos \theta,
\end{split}
\end{equation}
\begin{equation}
\begin{split}
\rho&\left\{ v_r \frac{\partial v_\theta}{\partial r} + \frac{v_\theta}{r} \frac{\partial v_\theta}{\partial \theta} + \frac{v_r v_\theta}{r} \right\} =\\ &-\frac{\partial \sigma_{r\theta}}{\partial r} - \frac{\sigma_{r\theta}}{r} - \frac{1}{r} \frac{\partial \sigma_{\theta\theta}}{\partial \theta} + \rho g \sin \theta.
\end{split}
\end{equation}

According to the  Mohr-Coulomb rheology \cite{jenike1959plastic}, the ratio of the principal stresses during flow is a constant ($K$), an extension of the Mohr-Coulomb condition at yield to a flowing system,
\begin{equation}
\frac{\sigma_1}{\sigma_2} = K,
\end{equation}
where $\sigma_1$ and $\sigma_2$ are the principal stresses. The stress ratio, $K$, is related to the angle of internal friction, $\beta$, by
\begin{equation}\label{eq:K}
K=\frac{1+f}{1- f}
\end{equation}
where $f=\sin\beta$. \citet{jenike1964} found that the angle of internal friction ($\beta$) in a flowing system was significantly larger than the value at yield. A second component of the rheological model is the coaxiality condition. The angle, $\gamma$, between the major principal axis for the stress and the $\theta$-axis is given by
\begin{equation}
\tan(2\gamma)=2\sigma_{r\theta}/(\sigma_{rr}-\sigma_{\theta\theta}).
\end{equation}
Similarly, the angle, $\gamma_D$, between the major principal axis for the rate of strain tensor, $\bm{D}$, and the $\theta$-axis, is given by
\begin{equation}
\tan(2\gamma_D)=2D_{r\theta}/(D_{rr}-D_{\theta\theta}),
\end{equation}
where $\bm{D}=(\nabla\bm{v}+\nabla\bm{v}^T)/2$. For coaxiality, the two angles must be equal, yielding the condition
\begin{equation}
\frac{D_{r\theta}}{D_{rr} - D_{\theta\theta}} = \frac{\sigma_{r\theta}}{\sigma_{rr}-\sigma_{\theta\theta}}.
\end{equation}

The boundary conditions for the system at the free surface ($r=R$) are
\begin{equation}
\sigma_{rr} = \sigma_{\theta\theta} = \sigma_{r\theta} = 0,
\end{equation}
and along the centerline ($\theta = 0$) are
\begin{equation}
    \frac{\partial \sigma_{rr}}{\partial \theta} = \frac{\partial \sigma_{\theta\theta}}{\partial \theta} = \sigma_{r\theta} = \frac{\partial v_r}{\partial \theta} = v_\theta = 0.
\end{equation}
The boundary conditions at the side wall of the hopper ($\theta = \theta_w$) are
\begin{equation}
\sigma_{r\theta} = -\mu_w \sigma_{\theta\theta}, \quad v_\theta = 0,
\end{equation}
where $\mu_w$ is the coefficient of wall friction. One more boundary condition is required at the exit of the hopper ($r=r_0$). \citet{savage1965mass} assumed the stress $\sigma_{rr}=0$ at the exit surface, which DEM simulation results given below, indicate is not valid. We thus do not apply any boundary condition at the exit and obtain the theoretical results in terms of an unknown constant, which is determined from the simulation results, as detailed below.

The same set of equations is valid for the $\mu(I)$ rheology \cite{jop2006}, but with a different definition of the stress ratio, $K$, as shown below. According to the $\mu(I)$ rheology \cite{jop2006} the effective friction coefficient is defined as
\begin{equation}
\mu(I)=\frac{\tau}{\sigma}
\end{equation}
where $\tau = (\sigma_1 - \sigma_2)/2$, and $\sigma = (\sigma_1 + \sigma_2)/2$, and $I = \dot{\gamma} d_p (\rho_p /\sigma)^{1/2}$ is the inertial number. Here $d_p$ and $\rho_p$ are the particle diameter and density, respectively. On simplifying, we get
\begin{equation}
\frac{\sigma_1}{\sigma_2} = \frac{1 + \mu}{1 - \mu} = K(I),
\end{equation}
which is identical to the  Mohr-Coulomb rheology \cite{jenike1959plastic}, but with the stress ratio dependent on the inertial number ($I$).

We rescale the governing equations using the following dimensionless variables: $\bar{r} = r/d_p$, $\bar{v}_i = v_i / (g d_p)^{1/2}$, $\bar{\sigma}_{ij} = \sigma_{ij} / \rho_p g d_p$, and $\bar{D}_{ij}=D_{ij}/(g/d_p)^{1/2}$. The dimensionless equations are
\begin{equation}\label{eq:cont}
\frac{\partial}{\partial \bar{r}} (\bar{r} \bar{v}_r) + \frac{\partial \bar{v}_\theta}{\partial \theta} = 0,
\end{equation}
\begin{equation}\label{eq:momr}
\begin{split}
\phi &\left\{ \bar{v}_r \frac{\partial \bar{v}_r}{\partial \bar{r}} + \frac{\bar{v}_\theta}{\bar{r}} \frac{\partial \bar{v}_r}{\partial \theta} - \frac{\bar{v}_\theta^2}{\bar{r}} \right\} =\\
&-\frac{\partial \bar{\sigma}_{rr}}{\partial \bar{r}} - \frac{1}{\bar{r}} \frac{\partial \bar{\sigma}_{r\theta}}{\partial \theta} + \frac{\bar{\sigma}_{\theta\theta} - \bar{\sigma}_{rr}}{\bar{r}} - \phi \cos \theta,
\end{split}
\end{equation}
\begin{equation}\label{eq:momt}
\begin{split}
\phi &\left\{ \bar{v}_r \frac{\partial \bar{v}_\theta}{\partial \bar{r}} + \frac{\bar{v}_\theta}{\bar{r}} \frac{\partial \bar{v}_\theta}{\partial \theta} + \frac{\bar{v}_r \bar{v}_\theta}{\bar{r}} \right\} =\\
&-\frac{\partial \bar{\sigma}_{r\theta}}{\partial \bar{r}} - \frac{\bar{\sigma}_{r\theta}}{\bar{r}} - \frac{1}{\bar{r}} \frac{\partial \bar{\sigma}_{\theta\theta}}{\partial \theta}  + \phi \sin \theta,
\end{split}
\end{equation}
using the relation $\rho=\rho_p\phi$, where $\phi$ is the solid volume fraction. The rescaled rheological equations are given by 
\begin{equation}
\bar{\sigma}_1 / \bar{\sigma}_2 = K,
\end{equation}
\begin{equation}\label{eq:coax}
\frac{\bar{D}_{r\theta}}{\bar{D}_{rr} - \bar{D}_{\theta\theta}} = \frac{\bar{\sigma}_{r\theta}}{\bar{\sigma}_{\theta\theta} - \bar{\sigma}_{rr}},
\end{equation}
and the boundary conditions in rescaled form are,
\begin{eqnarray}\label{eq:bcr2}
\bar{\sigma}_{rr}&=&\bar{\sigma}_{\theta\theta} = \bar{\sigma}_{r\theta} = 0,\quad\mbox{ at }\bar{r} = \bar{R},\\ \label{eq:bcth0}
\frac{\partial \bar{\sigma}_{rr}}{\partial \theta} &=& \frac{\partial \bar{\sigma}_{\theta\theta}}{\partial \theta} = \bar{\sigma}_{r\theta} = \frac{\partial \bar{v}_r}{\partial \theta} = \bar{v}_\theta = 0\quad\mbox{ at }\theta = 0,\\ \label{eq:bcthw}
\bar{\sigma}_{r\theta}&=&-\mu_w \bar{\sigma}_{\theta\theta},\quad \bar{v}_\theta = 0\quad\mbox{ at }\theta = \theta_w.
\end{eqnarray}
In what follows, we use dimensionless variables but drop the over bars for convenience.

An analytical solution of the above equations is not available, hence we present  a solution based on empirical scaling relations obtained from simulation data, in Sec.~\ref{sec:scth}. A perturbation solution for low wall friction ($\mu_w\ll1$) is given in the Supplemental Material. Predictions of both theories are compared to simulation data.

\section{Computational details}\label{sec:sim}
\begin{figure}
\begin{center}
\includegraphics[width=1.8in]{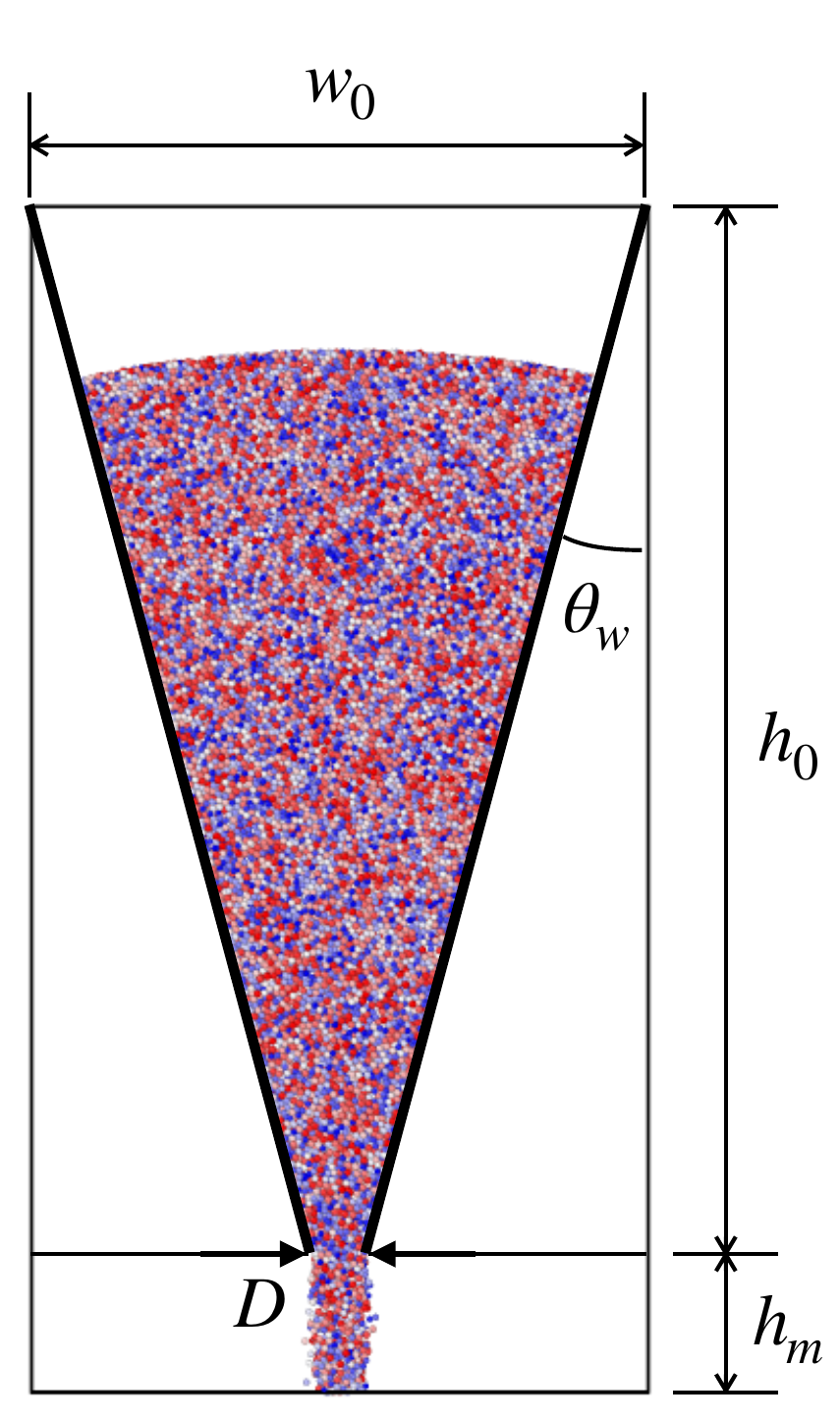}
\caption{Snapshot of the DEM simulation showing the system geometry, with the important dimensions indicated.}\label{fig:dem}
\end{center}
\end{figure}

A schematic view of the computational domain is shown in Fig.~\ref{fig:dem}, along with the important dimensions labelled. The side walls of the hopper, indicated by bold lines, are frictional. The base case values of the dimensions of the system are: $h_0=15$ cm, $h_m=2$ cm, $D=0.8$ cm, $\theta_w=15^{\circ}$, and $w_0=(D+2h_0\tan\theta_w)$. The thickness of the hopper normal to the plane of the flow ($z$-direction) is $b=1$ cm, and periodic boundary conditions are applied on the front and back surfaces. Discrete element method (DEM) simulations of the flow in the system are carried out using LAMMPS \cite{lammps}. The domain is extended below the exit plane to include the particles that have just exited the hopper, which exert a small back pressure.

About 60,000 particles of diameter $d_p = 1$ mm with a polydispersity of $\pm 10\%$, and with a density $\rho_p = 2.5$ g/cm$^3$ are used in the simulations. The normal and tangential deformations of the particles are assumed to follow the Hookean model with stiffnesses, $k_n$, and $k_t$, respectively. Viscous damping is taken to be proportional to the relative momentum normal to the contact and the damping coefficient is $\gamma_n$. The linear history model is used to compute the friction force, in which the relative tangential displacement, $\Delta \bm{s}$, is normalized to ensure that the tangential force ($k_t |\Delta \bm{s}|$) is limited by the maximum frictional force, $\mu_p |\bm{F}_n|$, where $\mu_p$ is the interparticle friction coefficient and $\bm{F}_n$ is the normal force. The same models are used for calculating forces between particles and walls, and the stiffness and damping parameters are the same as those for particle-particle contacts. The wall friction coefficient used is $\mu_w$. 

The values of the parameters used in the simulations are $k_n=2.568\times10^6$ dyn/cm, $k_t=2k_n/7$ and $\gamma_n=5.094\times10^3$ s$^{-1}$. The stiffness values ($k_n$, $k_t$) correspond to those used in the L3 model of \citet{silbert2001}. Simulations with a 100 fold larger stiffness values gave identical results \cite{yogi2026gravity}. The damping coefficient ($\gamma_n$) corresponds to a coefficient of restitution, $e=0.88$. The time step used is $dt=10^{-6}$ s, which is $2\%$ of the collision time for a pair of particles \cite{silbert2001}.

\begin{figure*}
\begin{center}
\includegraphics[width=5in]{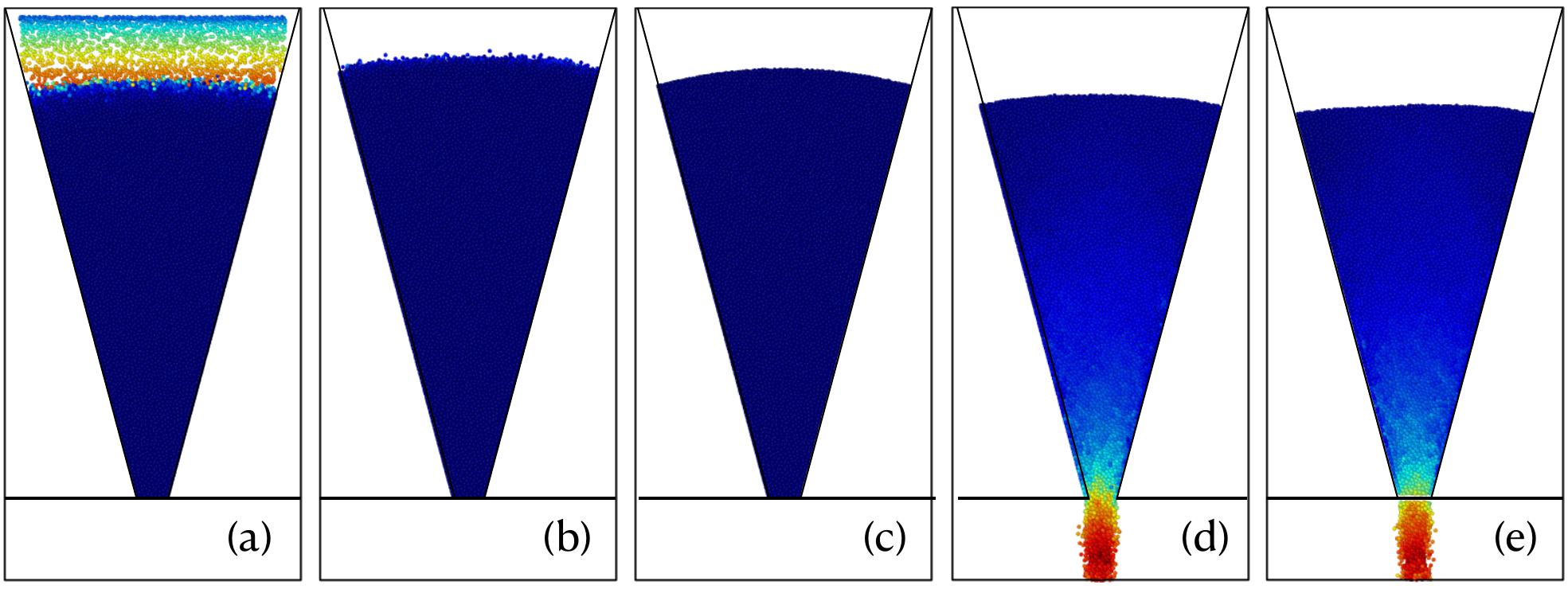}
\caption{Snapshots during the DEM simulation, showing the main steps. (a) Filling of the hopper. (b) Static bed after filling. (c) Static bed after shaping free surface. (d) Flow step after opening hopper exit. (e) State at the end of the flow step.}\label{fig:stp}
\end{center}
\end{figure*}

The computational procedure is devised to closely match the geometry used in the theoretical studies, and is shown in Fig.~\ref{fig:sch}, in which the top free surface is an arc of a circle with a specified radius. In the simulations, we shape the top free surface after the filling step and collect data for the static bed. We start the flow and after the flow has achieved a steady state, we shape the top surface again and collect data for a short time duration, to ensure that the change in height during this period is small (less than $5\%$). The process is repeated 50 times to obtain good averages. 

Snap shots of the process, highlighting the sequence of steps, are shown in Fig.~\ref{fig:stp}. In the first step, a sufficient number of particles are filled in the hopper (Fig.~\ref{fig:stp}(a)) to compensate for the particles that flow out of the system in the subsequent steps and that are removed in the shaping steps. After the particles settle to form a fixed bed (Fig.~\ref{fig:stp}(b)), the free surface is shaped by removing all particles with a radial position larger than a specified radius, $r_s=[r_d^2+5\tau\dot{m}_s/(\theta_w b\rho_p\phi)]^{1/2}$ cm, (Fig.~\ref{fig:stp}(c)), and the data for the static bed are collected at this stage. Here $r_d=(h_0+r_0-2.6)$ cm, $\dot{m}_s$ is the steady state flow rate and $5\tau$ is the time for the flow to achieve a steady state. $r_s$ is the free surface radius such that the the free surface radius is $r_d$ after a flow of duration $5\tau$. Here $\tau$ is the characteristic time for the time-variation of the mass flow rate, which is found to follow an exponential rise of the form
\begin{equation}\label{eq:mdt}
\dot{m}_t=\dot{m}_s\left[1-\exp\left(-t_f/\tau\right)\right],
\end{equation}
where $t_f$ is the time after which the flow is started. The exit is opened for a specified duration ($t=5\tau$), after which the flow is steady, and the free surface is shaped again, removing all particles with radius larger then $r_d$. During steady state flow ($t_f>5\tau$) data are collected for a short duration, during which the change in height is $\Delta r=0.5$ cm (Fig.~\ref{fig:stp}(d)). The exit is then closed (Fig.~\ref{fig:stp}(e)), and the process is repeated.

The data are averaged over the time durations of collection, which were $0.1$ s for the static bed and about $0.1$ s for the flow. These data are further averaged over 50 runs and mean values and standard errors are reported below. The system variables studied are the solid volume fraction and stresses for the static system and the volume fraction, velocities and stresses for the flowing system. The spatial distributions were studied by binning the data using a grid in cylindrical coordinates with bins of dimension $\Delta r \times \Delta \theta =2\mbox{ mm }\times1^\circ$.

\begin{table}
\begin{center}
\caption{Hopper parameters ($D$, $\theta_w$, $h_0$ and $\mu_w$, Fig.~\ref{fig:sch}) and particle properties ($\mu_p$ and $d_p$) varied in the simulations. All variables are rescaled except for the particle diameter ($d_p$) and wedge angle ($\theta_w$).\label{tab:par}}
\begin{tabular}{|c|c|}\hline
Parameter & Values \\ \hline\hline
$d_p$ (mm)& 1, 1.5, 2 \\
$\mu_p$ & 0, 0.05, 0.1, 0.2, 0.3, 0.4, 0.5, 0.6 \\
$\mu_w$ & 0, 0.05, 0.1, 0.2, 0.3, 0.4, 0.5, 0.6 \\
$h_0$ & 120, 150, 180 \\
$D$ & 6, 8, 10, 12, 14 \\
$\theta_w$ (deg.) & 10, 15, 20, 25 \\ \hline
\end{tabular}
\end{center}
\end{table}

The base case parameters for the system in rescaled form, except for $d_p$ and $\theta_w$, are as follows: $d_p=1$ mm, $\mu_p=0.5$, $\mu_w=0.5$, $h_0=15$, $\theta_w=15^\circ$, and $D=8$. The parameters are varied one at a time, keeping the remaining parameters equal to the base case values. The values of the parameters used in the computations are given in Table~\ref{tab:par}.

\section{Results and discussion}\label{sec:res}

In this section, we first present results for the base case (Sec.~\ref{sec:bc}), to show the typical behaviour of the system. Scaling relations for the velocity and stress fields for the base case are presented next (Sec.~\ref{sec:sca}), together with a discussion of the rheology of the flow. A theory based on the scaling relations is given in Sec.~\ref{sec:scth}, and predictions of the theory are compared to the simulation results. The results of the parametric study are presented in Sec.~\ref{sec:par}, and the variation of the mass flow rate with system parameters is dicussed in Sec.~\ref{sec:md}.  The larger than gravity acceleration of particles in the exit region of the hopper as briefly discussed in Sec.~\ref{sec:acc}. All results are in terms of dimensionless variables, unless specifically indicated.

\subsection{Base case}\label{sec:bc}
\begin{figure}
\begin{center}
\includegraphics[width=\fw in]{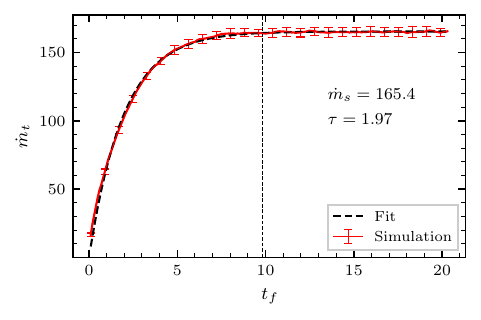}
\caption{Variation of the mass flow rate ($\dot{m}$) with time ($t_f$) for the base case ($D=8$, $\mu_w=0.5$, $\mu_p=0.5$, $h=120$, $\theta_w=15$ deg.) The dashed line is a fit of Eq.~(\ref{eq:mdt}), with the fitted parameters shown in the figure. The vertical dotted line shows the start of the interval for averaging ($t=5\tau$).}\label{fig:mdt}
\end{center}
\end{figure}

Fig.~\ref{fig:mdt} shows the variation of the mass flow rate ($\dot{m}_t$) with time of flow ($t_f$). The flow rate increases exponentially with time and achieves a steady state value ($\dot{m}_s$). The dashed line is a fit of Eq.~(\ref{eq:mdt}) to the data, and a very good match is obtained. The values of the fitted parameters are shown in the figure. The duration to achieve a steady state, $t_f=5\tau$, corresponding to a dimensional time of about 0.1 s, is shown by the vertical dotted line. The mass flow rate is constant beyond $t_f=5\tau$, and the averaging of the data is done for a duration $\Delta t_f\approx10$ for $t_f>5\tau$.

\begin{figure}
\begin{center}
\includegraphics[width=\fw in]{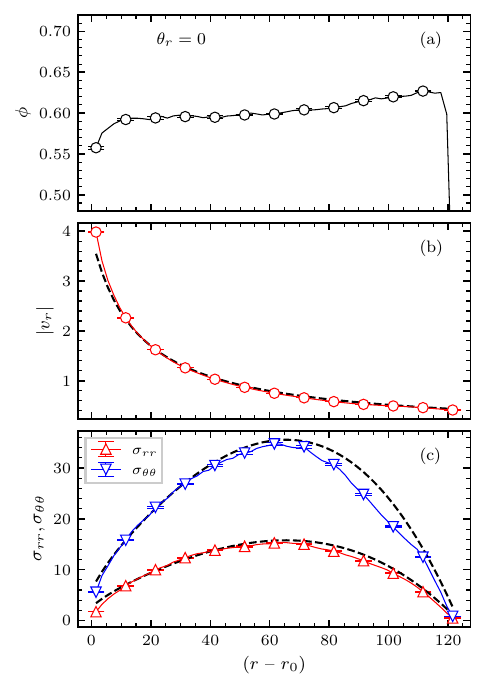}
\caption{Variation of (a) the solid volume fraction ($\phi$), (b) magnitude of the radial velocity ($|v_r|$), (c) radial normal stress ($\sigma_{rr}$) and tangential normal stress ($\sigma_{\theta\theta}$) with radial distance from the exit ($r-r_0$), along the centerline of the hopper ($\theta_r=0$) for the base case. The error bars indicate the standard error. The dashed lines are predictions of the theory.}\label{fig:bcr}
\end{center}
\end{figure}

Consider first, the results for the flowing system. Fig.~\ref{fig:bcr} shows the variation of the solid volume fraction ($\phi$), the magnitude of the radial velocity ($|v_r|$) and the normal stresses ($\sigma_{rr}, \sigma_{\theta\theta}$) with radial distance from the exit ($r-r_0$), along the centerline of the hopper ($\theta=0$). The error bars show the standard error, which is small for all the data in the figure. The solid fraction increases slightly with radial distance, and decreases sharply near the exit and the free surface. The velocity decreases significantly with radial distance from the exit ($r-r_0$), and the velocity is highest at the exit ($r=r_0$). Both normal stresses are zero at the free surface and are small but non-zero at the exit, with a maximum near the middle of the hopper. The non-zero exit stresses are due to the confining pressure of the side walls, as well as the back pressure of the particles that have just exited the hopper. The tangential normal stress ($\sigma_{\theta\theta}$) is approximately twice as large the radial normal stress ($\sigma_{rr}$). The shear stress ($\sigma_{r\theta}$) is zero along the centerline, due to symmetry, and is hence not shown.

\begin{figure}
\begin{center}
\includegraphics[width=\fw in]{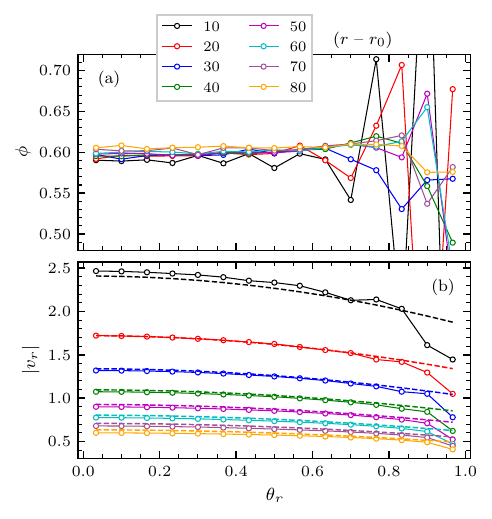}
\caption{Variation of (a) the solid volume fraction ($\phi$) and (b) the magnitude of the radial velocity ($|v_r|$) with scaled angle ($\theta_r$) at different radial distances from the exit ($r-r_0$), for the base case. The dashed lines are predictions of the theory.}\label{fig:phvr}
\end{center}
\end{figure}

The variation of the solid fraction ($\phi$) and the magnitude of the radial velocity ($|v_r|$), with scaled angle, $\theta_r=\theta/\theta_w$, for different radial distances from the exit ($r-r_0$) given in the legend, is shown in Fig.~\ref{fig:phvr}. The solid volume fraction is nearly constant with angle, except for fluctuations near the walls, due to layering of particles (Fig.~\ref{fig:phvr}(a)). The velocity appears to vary quadratically with angle at each radial position (Fig.~\ref{fig:phvr}(b)), and this is examined in more detail in Sec.~\ref{sec:sca}. 

\begin{figure}
\begin{center}
\includegraphics[width=\fw in]{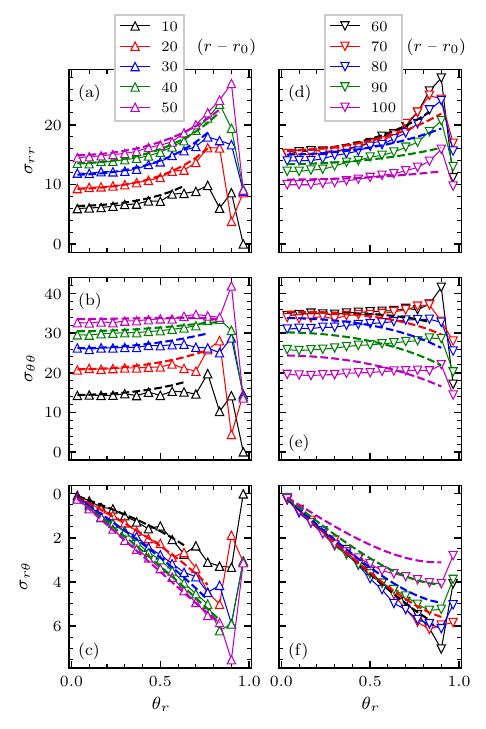}
\caption{Variation of the stresses ($\sigma_{rr}$, $\sigma_{\theta\theta}$, $\sigma_{r\theta}$) with scaled angle ($\theta_r$) at different radial distances from the exit ($r-r_0$), for the base case. The figures in the left column ((a), (b), (c)) correspond to increasing normal stresses with radial distance, and the figures in the right column ((d), (e), (f)) to decreasing normal stresses with radial distance. The dashed lines are predictions of the theory.}\label{fig:st}
\end{center}
\end{figure}

Fig.~\ref{fig:st} shows the variation of the stresses  ($\sigma_{rr}$, $\sigma_{\theta\theta}$ and $\sigma_{r\theta}$) with the scaled angle, $\theta_r$, at different radial positions, ($r-r_0$). The left column corresponds to radial positions below the stress maximum and the right column to positions above the stress maximum. The radial normal stress ($\sigma_{rr}$) is minimum at the centerline ($\theta_r = 0$) and increases significantly with the scaled angle ($\theta_r$), for all radial positions. The tangential normal stress $(\sigma_{\theta \theta})$ is nearly constant at a fixed radial distance, and increases only slightly with $\theta_r$. In contrast, the shear stress $(\sigma_{r \theta})$ is zero at the centerline and decreases linearly with increasing $\theta_r$ for most of the radial positions. Deviations from linearity occur near the exit and near the  free surface, for $(r - r_0) = 10$ and $90,100$. The maximum shear stress is about five-fold smaller in magnitude than the tangential normal stress at the same radial position, despite the relatively high wall friction coefficient ($\mu_w=0.5$). 

\begin{figure}
\begin{center}
\includegraphics[width=\fw in]{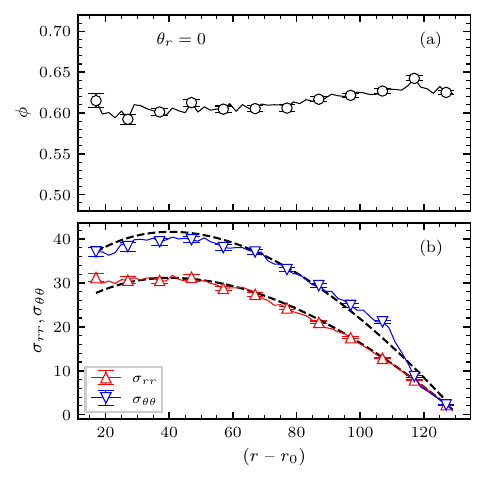}
\caption{Variation of (a) the solid volume fraction ($\phi$), (b) radial normal stress ($\sigma_{rr}$) and tangential normal stress ($\sigma_{\theta\theta}$) with radial distance from the exit ($r-r_0$), along the centerline of the hopper ($\theta_r=0$), for the static base case. The error bars indicate the standard error. The dashed lines are predictions of the theory.}\label{fig:bcrs}
\end{center}
\end{figure}

Consider the results for the static case next. Fig.~\ref{fig:bcrs} shows the variation of the solid fraction ($\phi$) and normal stresses ($\sigma_{rr}$, $\sigma_{\theta\theta}$) with radial distance from the base ($r-r_0$), along the centerline of the hopper ($\theta_r=0$). The solid fraction is nearly constant, and close to that for the flowing system (Fig.~\ref{fig:bcr}(a)). The normal stresses are zero at the free surface and are of significant magnitude at the base ($r=r_0$), with a shallow maximum. The values of the maximum stress for each of the normal stresses in the static system are larger than the corresponding stresses in the flowing system.

\begin{figure}
\begin{center}
\includegraphics[width=\fw in]{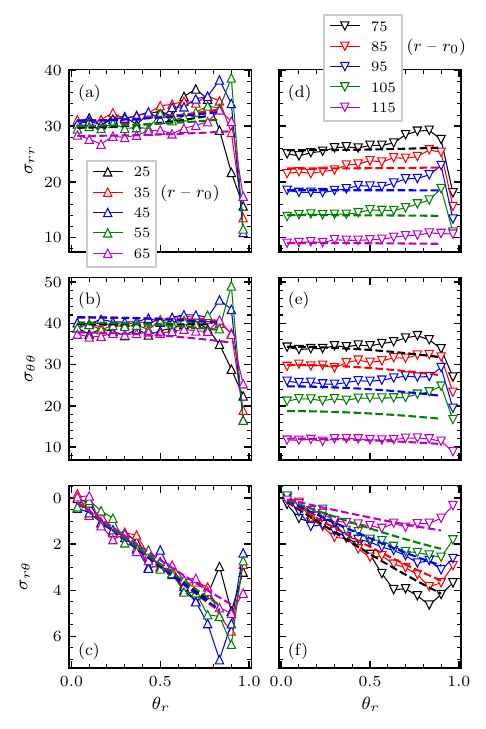}
\caption{Variation of the stresses ($\sigma_{rr}$, $\sigma_{\theta\theta}$, $\sigma_{r\theta}$) with scaled angle ($\theta_r$) at different radial distances from the exit ($r-r_0$), for the static base case. The figures in the left column ((a), (b), (c)) correspond to increasing normal stresses with radial distance, and the figures in the right column ((d), (e), (f)) to decreasing normal stresses with radial distance. The dashed lines are predictions of the theory.}\label{fig:sts}
\end{center}
\end{figure}

The variation of the stresses ($\sigma_{rr}$, $\sigma_{\theta\theta}$, $\sigma_{r\theta}$), in the static system, with the scaled angle ($\theta_r$) at different radial positions ($r-r_0$) is shown in Fig.~\ref{fig:sts}. For a given radial position, the radial normal stress ($\sigma_{rr}$) is lowest at the centerline ($\theta_r=0$) and increases with angle, $\theta_r$, while the tangential normal stress ($\sigma_{\theta\theta}$) is nearly constant with angle. The shear stress is zero at the centerline ($\theta_r=0$), and its magnitude increases linearly with scaled angle. These results are qualitatively similar to those for the flowing system. The magnitudes of the maximum radial normal stress and tangential normal stress are larger than those for the flowing system, as noted above, however, the magnitudes of the shear stresses in the flowing and static systems are nearly the same. We obtain scaling relations for the velocity and stresses next.

\begin{figure}
\begin{center}
\includegraphics[width=\fw in]{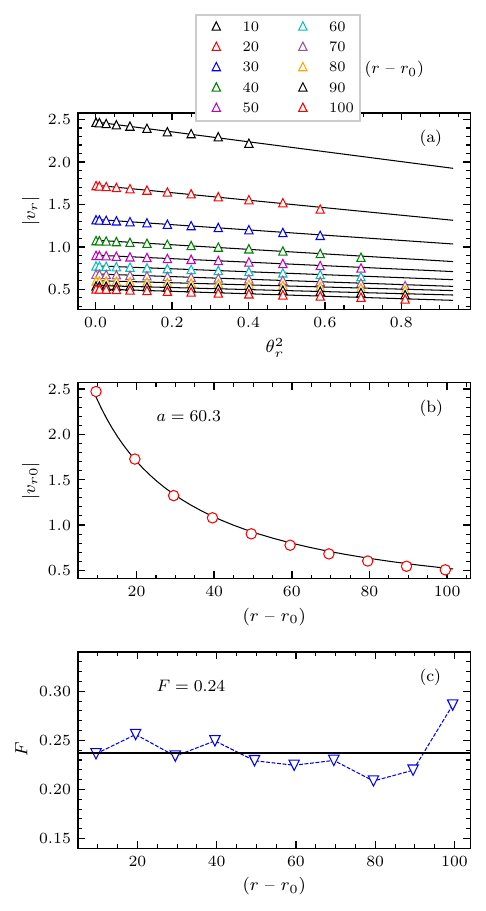}
\caption{(a) Variation of the magnitude of the radial velocity ($|v_r|$) with the square of the scaled angle ($\theta_r^2$) for the base case (symbols), and fits of Eq.~(\ref{eq:vrfit}) to the data (lines). Variation of the fitted values of (b) the centerline velocity ($v_{r0}$ and (c) the velocity slope parameter ($F$), with radial distance from the exit ($r-r_0$) (symbols) obtained from (a). The line in (b) is a fit of Eq.~(\ref{eq:vr0}) and in (c) a fit of a constant value, with the values of the fitted parameters ($a$, $F$) indicated in the respective figures.}\label{fig:vrfit}
\end{center}
\end{figure}

\subsection{Scaling relations and rheology}\label{sec:sca}
We consider the scaling of the radial velocity ($v_r$) based on the form obtained by \citet{brennen1978} from the perturbation analysis for small wall friction ($\mu_w\ll1$), given by
\begin{equation}\label{eq:vrfit}
v_r = -v_{r0} (1 - F \theta_r^2),
\end{equation}
with 
\begin{equation}\label{eq:vr0}
v_{r0} = a/r,
\end{equation}
where $a$ and $F$ are constants. \citet{yogi2026granular} showed the validity of this scaling, even for large wall friction coeffcients ($\mu_w=0.4$), by means of experiments in a quasi-2D wedge shaped hopper and DEM simulations in quasi-2D and 3D wedge-shaped hoppers.

Fig.~\ref{fig:vrfit}(a) shows the variation of the magnitude of the radial velocity ($|v_{r}|$) with the square of the scaled angle $(\theta_r^2)$, at different radial distances from the exit ($r-r_0$).  Data within a distance $2d_p$ from the wall are omitted, to omit the effects of particle layering near the wall ($\theta_r=1$). The lines are fits of Eq.~(\ref{eq:vrfit}) at each radial position, and good fits are obtained for all radial positions. The fitted values of the centerline velocity, $v_{r0}$, and the slope of the normalized velocity profile ($|v_r|/v_{r0}$), $F$, at different radial positions are shown in Figs.~\ref{fig:vrfit}(b) and (c), respectively. The centerline velocity varies as in Eq.~(\ref{eq:vr0}), and a good fit is obtained for $a = 60.3$ (Fig.~\ref{fig:vrfit}(b)). The slope, $F$, is nearly constant and the solid line in Fig.~\ref{fig:vrfit}(c) is a least squares fit to the data omitting the last two points ($r>80$).

\begin{figure}
\begin{center}
\includegraphics[width=\fw in]{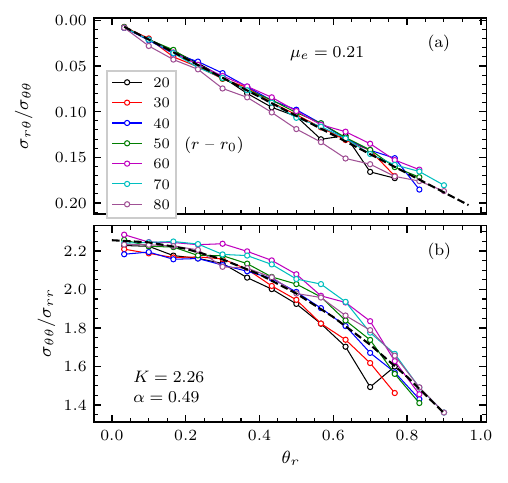}
\caption{Variation of the ratio of (a) the shear stress to the tangential normal stress ($|\sigma_{r\theta}|/\sigma_{\theta\theta}$) and (b) the tangential normal stress to the radial normal stress ($\sigma_{\theta\theta}/\sigma_{rr}$), with the scaled angle ($\theta_r$), for the base case. The lines are fits of Eqs.~(\ref{eq:scrt}) and (\ref{eq:sctt}) and the fitted values of the parameters are given in the respective figures.}\label{fig:srat}
\end{center}
\end{figure}

We consider scaling of the stress components next. The stresses may be written in terms of the Sokolovski variables \cite{sokolovski1960statics} for a material that follows the  Mohr-Coulomb rheology \cite{jenike1959plastic} as
\begin{eqnarray}
\sigma_{\theta\theta}&=&\sigma(1+f\cos2\gamma),\\
\sigma_{rr}&=&\sigma(1-f\cos2\gamma),\\
\sigma_{r\theta}&=&-\sigma f\sin2\gamma,
\end{eqnarray}
where $\sigma=(\sigma_{rr}+\sigma_{\theta\theta})/2$ is the mean stress. For low wall friction ($\mu_w$) the shear stress ($\sigma_{r\theta}$) is small \cite{brennen1978}, and hence the angle $\gamma$ is also small. Further, we assume, based on the analysis of \citet{brennen1978}, that $\gamma=C_{\gamma}\theta_r$, where $C_{\gamma}$ is a constant. Expanding the above equations for small $\gamma$, keeping only terms that are larger than $O(\gamma^2)$, and using the expression for $\gamma$, we obtain
\begin{eqnarray}\label{eq:scrt}
\frac{\sigma_{r\theta}}{\sigma_{\theta\theta}}&=&-\frac{2C_{\gamma}f}{1+f}\theta_r=-\mu_e\theta_r,\\ \label{eq:sctt}
\frac{\sigma_{\theta\theta}}{\sigma_{rr}}&=&\frac{1+f}{1-f}\left(1-\frac{4C_{\gamma}^2}{(1-f)^2}\theta_r^2\right)=K(1-\alpha\theta_r^2),
\end{eqnarray}
using Eq.~(\ref{eq:K}). Further, the wall boundary condition, Eq.~(\ref{eq:bcthw}), implies that $\mu_e=\mu_w$, which gives  $C_{\gamma}=\mu_w(1+f)/2f$, and
\begin{equation}\label{eq:al}
\alpha=\mu_w^2K(K+1)/(K-1).
\end{equation}
Thus, the stress ratios are obtained in terms of two parameters, the wall friction coefficient, $\mu_w$, and the internal friction coefficient, $f$. Although the above scaling results are valid for low wall friction, we use them for high wall friction as well, treating $\mu_e$, $K$ and $\alpha$ as empirical constants. The same scaling relations are also valid for the static system, as well as the $\mu(I)$ rheology \cite{jop2006}. In the latter, $f$ is replaced by $\mu(I)$, and thus the parameters vary with the inertial number, $I$. We compare the predictions of the scaling relations to simulation data next.

Fig.~\ref{fig:srat}(a) shows the variation of the ratio of the  shear stress to the tangential normal stress $(\sigma_{r \theta} / \sigma_{\theta \theta})$ with the scaled angle ($\theta_r$). The data for different radial positions in the middle region of the hopper, $(r - r_0) \in (20, 80)$, collapse to a single straight line, with a slope $\mu_e$, indicated in the figure. Thus, the scaling given in Eq.~(\ref{eq:scrt}) is valid, and the effective wall friction coefficient, $\mu_e$, is independent of the radial position. However, $\mu_e$ is significantly lower than the wall friction coefficient, $\mu_w=0.5$. This is because the relative motion at the particle-wall contact is not purely due to sliding but may also involve rolling, which has a lower friction coefficient \cite{lin2023modelling}.

Fig.~\ref{fig:srat}(b) shows the variation of the ratio of the tangential normal stress to the radial normal stress ($\sigma_{\theta\theta}/\sigma_{rr}$) with the scaled angle ($\theta_r$). Data for the different radial positions again collapse to a single line, though the scatter in the data is larger in this case compared to the shear stress scaling (Fig.~\ref{fig:srat}(a)). The dashed line in the figure is a fit of Eq.~(\ref{eq:sctt}) to all the data, and the fitted parameters are indicated in the figure. 

The fits shown in Figs.~\ref{fig:srat}(a) and (b) indicate that the scaling relations given in Eqs.~(\ref{eq:scrt}) and (\ref{eq:sctt}) give a good description of the data. Further, the rheology is friction dominated, since the scaling relations are independent of the radial distance, over which the inertial number, $I$, varies significantly, as shown below. We consider next the validity of the  Mohr-Coulomb rheology \cite{jenike1959plastic}, and the coaxiality condition using the DEM simulation data and the fits of Eqs.~(\ref{eq:scrt}) and (\ref{eq:sctt}).

\begin{figure}
\begin{center}
\includegraphics[width=\fw in]{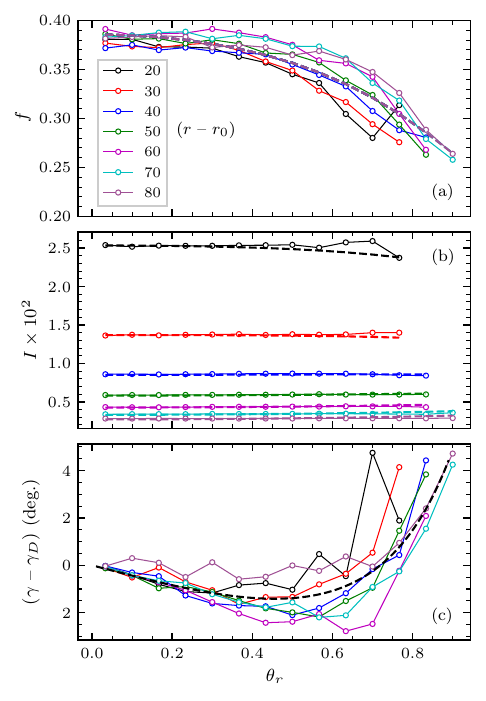}
\caption{Variation of (a) the internal friction coefficient ($f$), (b) the inertial number ($I$), and (c) the ratio of the tangents of the principal angles of the stress and strain rate ($\tan(2\gamma)/\tan(2\gamma_D)$), with the scaled angle ($\theta_r$), for the base case. The lines are computed from the scaling relations.}\label{fig:fgm}
\end{center}
\end{figure}

Figs.~\ref{fig:fgm}(a) and (b) show the variation of the internal friction coefficient, $f$, and the inertial number ($I$), with scaled angle ($\theta_r$) for different radial positions. The internal friction coefficient is computed as $f=\tau/\sigma$, where $\tau=[(\sigma_{\theta\theta}-\sigma_{rr})^2+4\sigma_{r\theta}^2]^{1/2}/2$. Using the scaling relations (Eqs.~(\ref{eq:scrt}) and (\ref{eq:sctt})), and simplifying we also obtain 
\begin{equation}\label{eq:fa}
f = [(\kappa-1)^2+4\mu_e^2\kappa^2]^{1/2}/(\kappa+1),
\end{equation}
where $\kappa=K(1-\alpha\theta_r^2)$. In Fig.~\ref{fig:fgm}(a), the symbols are the simulation data and the dashed line is computed from Eq.~(\ref{eq:fa}), using fitted values of $\mu_e$, $K$ and $\alpha$. The internal friction coefficient ($f$) is independent of the radial position, and thus the inertial number, $I$, at each $\theta_r$, as required by the Mohr-Coulomb rheology \cite{jenike1959plastic}. The result is reasonable considering the low values of the inertial number ($I<0.025$, Fig.~\ref{fig:fgm}(b)). However, the internal friction coefficient, $f$, decreases with increasing angle ($\theta_r$), despite the inertial number ($I$) being constant with $\theta_r$ at each radial position (Fig.~\ref{fig:fgm}(b)). The results indicate that the Mohr-Coulomb rheology \cite{jenike1959plastic} is valid in the region considered, but the friction coefficient reduces with angle, $\theta_r$. This appears to be an effect of the shear component of the velocity gradient, $D_{r\theta}$, which increases with angle, $\theta_r$, and causes a reduction in local internal friction. The reduction in the friction coefficient is significant (up to 30\%), and needs to be accounted for. These results are in contrast to those for frictionless walls \cite{momin2025granular}, in which $D_{r\theta}=0$ and the rheology is well-described by the Mohr-Coulomb rheology \cite{jenike1959plastic} with a constant internal friction coefficient, $f$, in the bulk and by the $\mu(I)$ rheology \cite{jop2006} in the exit region. We carried out a similar analysis, as above, for the present system in the exit region, where $I\sim0.1$, and obtained similar scaling for the stress ratios. The parameters in this case are dependent on the inertial number, $I$. However, the $\mu$ versus $I$ data do not collapse to a single curve.

Fig.~\ref{fig:fgm}(c) shows the variation of the difference in the angle of the principal stress axis ($\gamma$) and the angle of the principal rate of strain axis ($\gamma_D$) with scaled angle ($\theta_r$). Results obtained using the fitted scaling relations (dashed lines). and the simulation results (symbols), are shown. The difference is small ($|\theta-\theta_D|<2^{\circ}$) for $\theta_r<0.8$, and increases to about 4$^{\circ}$ at $\theta_r\approx1$. Thus, the coaxiality condition is approximately satisfied at low angles ($\theta_r<0.8$), but there are significant deviations from coaxiality near the wall ($\theta_r>0.9$).

\begin{figure}
\begin{center}
\includegraphics[width=\fw in]{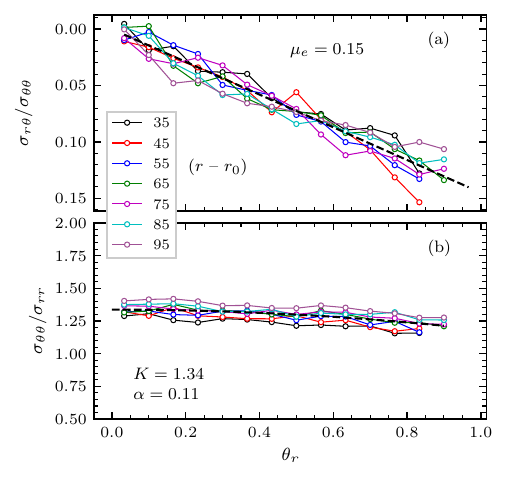}
\caption{Variation of the ratio of (a) the shear stress to the tangential normal stress ($\sigma_{r\theta}/\sigma_{\theta\theta}$) and (b) the tangential normal stress to the radial normal stress ($\sigma_{\theta\theta}/\sigma_{rr}$), with the scaled angle ($\theta_r$), for the static base case. The lines are fits of Eqs.~(\ref{eq:scrt}) and (\ref{eq:sctt}) and the fitted values of the parameters are given in the respective figures.}\label{fig:srats}
\end{center}
\end{figure}

Consider the scaling of data for the static systems next. Scaling relations for the stress ratios, $\sigma_{r\theta}/\sigma_{\theta\theta}$ and  $\sigma_{\theta\theta}/\sigma_{rr}$, with the scaled angle, $\theta_r$ are shown in Fig.~\ref{fig:srats}. The results indicate that scaling relations, given in Eqs.~(\ref{eq:scrt}) and (\ref{eq:sctt}) are also valid for the static case. However, the values of the parameters, $\mu_e$, $K$ and $\alpha$, are significantly smaller for the static system relative to the flowing one.

\subsection{Theory based on the scaling relations}\label{sec:scth}
The perturbation solution, given in the Supplemental Material, gives predictions that deviate significantly from the simulation results for wall friction coefficients that are not small ($\mu_w\ge0.1$). The Mohr-Coulomb rheology \cite{jenike1959plastic} and the $\mu(I)$ rheology \cite{jop2006} also do not apply directly to the flow, as shown above. Further, coaxiality condition is violated near the wall. Consequently, we utilize the empirical scaling relations, obtained above, for the velocity and stress ratios, in place of the constitutive equation and the coaxiality condition, to obtain the velocity and stress distribution in the hopper.

The scaling relations used in the analysis are
\begin{eqnarray}\label{eq:vra}
v_{r} &=& -A(\theta)/r, \quad v_{\theta} = 0,\\
\label{eq:srta}
\sigma_{r\theta} &=& -\mu_{e}\sigma_{\theta\theta}\theta_{r},\\
\label{eq:stta}
\sigma_{\theta\theta} &=& \kappa(\theta)\sigma_{rr},
\end{eqnarray}
where $A = a(1 - F\theta_{r}^{2})$.

The continuity equation (Eq.~(\ref{eq:cont}))) is identically satisfied by the velocity field given in Eq.~(\ref{eq:vra}). Substituting for $v_r$, $v_{\theta}$, $\sigma_{r\theta}$ and $\sigma_{\theta\theta}$, using Eqs.~(\ref{eq:vra})-(\ref{eq:stta}), in the $r$-momentum balance equation (Eq.~(\ref{eq:momr})), and simplifying, we get
\begin{equation}\label{eq:dsrr1}
\begin{split}
\frac{\partial\sigma_{rr}}{\partial r}& - \left[\kappa(1 + \mu_{e}/\theta_{w}) - 1\right]\frac{\sigma_{rr}}{r} =\\ &\frac{\phi A^{2}}{r^{3}} - \phi g \cos\theta + \frac{\mu_{e}\theta_{r}}{r}\frac{\partial\sigma_{\theta\theta}}{\partial\theta}.
\end{split}
\end{equation}
Neglecting the last term in Eq.~(\ref{eq:dsrr1}), since $\sigma_{\theta\theta}$ is nearly constant with respect to $\theta$, we obtain
\begin{equation}\label{eq:sava}
\frac{\partial\sigma_{rr}}{\partial r} - p\frac{\sigma_{rr}}{r} = \frac{\phi A^{2}}{r^{3}} - \phi g'
\end{equation}
where
\begin{equation}
p = {\kappa(1 + \mu_{e}/\theta_{w}) - 1},
\end{equation}
and $g'=\cos\theta$. Including the $\partial\sigma_{\theta\theta}/\partial\theta$ term in the solution is straightforward, however, this does not make a significant difference in the predictions of the model. Eq.~(\ref{eq:sava}) is identical to the \citet{savage1965mass} equation, but with $p$, $A$ and $g'$ being known functions of $\theta$, and has the solution
\begin{equation}\label{eq:srra}
\begin{split}
\sigma_{rr}(r,\theta)&=\frac{\phi g'r}{p-1}\left[1-\left(\frac{r}{R}\right)^{p-1}\right]\\&-\frac{\phi A^2}{(p+2)r^2}\left[1-\left(\frac{r}{R}\right)^{p+2}\right],
\end{split}
\end{equation}
using the boundary condition given in Eq.~(\ref{eq:bcr2}), $\sigma_{rr}=0$ at $r=R$. The other stress components are given by Eqs.~(\ref{eq:srta}) and (\ref{eq:stta}), and the velocity by Eq.~(\ref{eq:vra}). The same equations are valid for the static case as well, but with $A=0$.

We compare predictions of the theory to computational results next. The dashed lines in Figs.~\ref{fig:bcr}(b) and (c) are predictions of Eqs.~(\ref{eq:vra}), (\ref{eq:stta}) and (\ref{eq:srra}) for the radial velocity ($|v_r|$), the radial normal stress ($\sigma_{rr}$) and the tangential normal stress ($\sigma_{\theta\theta}$) along the centerline of the hopper ($\theta_r=0$), using fitted values of the constants $a$, $\mu_e$ and $K$, and taking the solid volume to be $\phi=0.6$. The agreement between theory and computational results is very good except for small deviations in the tangential normal stress predictions near the upper part of the hopper ($(r-r_0)\approx90$). 

The predictions of Eq.~(\ref{eq:vra}) for the radial velocity ($|v_r|$) variation with angle ($\theta_r$) at different radial positions ($r-r_0$), using fitted values of the parameters $a$ and $F$, (dashed lines) are compared to the simulation results (symbols) for the base case in Fig.~\ref{fig:phvr}(b) . The agreement between the two is again very good. Similarly, the predictions of Eqs.~(\ref{eq:srta}), (\ref{eq:stta}) and (\ref{eq:srra}) for the variation of the radial normal stress ($\sigma_{rr}$), the tangential normal stress ($\sigma_{\theta\theta}$) and the shear stress ($\sigma_{r\theta}$) (dashed lines) with scaled angle ($\theta_r$) are compared to simulation results (symbols) in Fig.~\ref{fig:st}. The agreement between theory and experiment is good for $(r-r_0)\le80$ for all the stresses, but deviations are seen for larger radial positions. In particular, the predicted tangential normal stress ($\sigma_{\theta\theta}$) shows a decrease with angle for $(r-r_0)>50$, in contrast to the simulation results which show a small increase. 

A comparison of the theory (dashed lines) to simulation results (symbols) for the static system for the normal stresses ($\sigma_{rr}$, $\sigma_{\theta\theta}$) along the centerline of the hopper ($\theta_r=0$) is shown in Fig.~\ref{fig:bcrs}. Predictions of the theory (dashed lines) for the normal stresses ($\sigma_{rr}$, $\sigma_{\theta\theta}$) and shear stress ($\sigma_{r\theta}$) with angle ($\theta_r$) at different radial positions ($r-r_0$) are compared to the simulation results (symbols). Fitted values of the constants $\mu_e$, $K$ and $\alpha$ are used. The agreement between the theory and simulations is again good, except for the predictions of the tangential normal stress ($\sigma_{\theta\theta}$) at the higher radial positions.

Thus, predictions of the theory are quite good except for deviations in the tangential normal stress at higher radial positions. These deviations may be due to boundary effects near the free surface. The scaling relations also deviate from the data in this region.

\subsection{Results of the parametric study}\label{sec:par}
We summarize briefly, the behaviour of the flow with variation of parameters. The results for varying hopper parameters (orifice width, $D$, wedge angle, $\theta_w$, and height of the hopper $h_0$) are presented first, followed by results for varying particle properties (particle diameter, $d_p$, and particle friction coefficient $\mu_w$). For brevity, only profiles along the centerline of the hopper ($\theta_r=0$) are presented in each case. The variation with angle, $\theta_r$, is qualitatively similar to that for the base case, shown in the previous section, and is hence not given here. 

\begin{figure}
\begin{center}
\includegraphics[width=\fw in]{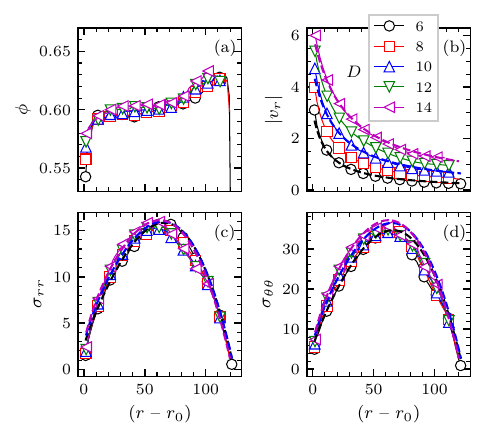}
\caption{Variation of (a) the solid volume fraction ($\phi$), (b) magnitude of the radial velocity ($|v_r|$), (c) radial normal stress ($\sigma_{rr}$) and (d) tangential normal stress ($\sigma_{\theta\theta}$) with radial distance from the exit ($r-r_0$), along the centerline of the hopper ($\theta_r=0$) for different orifice widths ($D$). The dashed lines are predictions of the theory.}\label{fig:Dr}
\end{center}
\end{figure}

Consider first the effect of increasing \textit{orifice width}, $D$. Fig.~\ref{fig:Dr} shows the variation of the solid volume fraction ($\phi$), magnitude of the radial velocity ($|v_r|$), the radial normal stress ($\sigma_{rr}$) and the tangential normal stress ($\sigma_{\theta\theta}$) with radial distance from the exit ($r-r_0$), along the centerline of the hopper ($\theta = 0$), for different orifice widths ($D$).  The radial velocity at the exit ($r = r_0$) increases significantly with orifice width, from $|v_r| \approx 3$ for a width $D = 6$ to $|v_r| \approx 6$ for $D = 14$. Despite the near doubling of the velocity, the solid fraction ($\phi$) and the normal stresses ($\sigma_{rr}, \sigma_{\theta\theta}$) are nearly unchanged with increase in the orifice width. The nearly constant stresses with increase in velocity implies that the flow is dominated by friction forces, which are independent of the relative velocity between contacting surfaces.

\begin{figure}
\begin{center}
\includegraphics[width=\fw in]{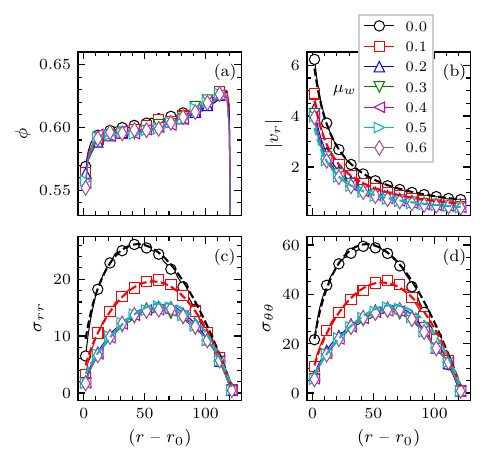}
\caption{Variation of (a) the solid volume fraction ($\phi$), (b) magnitude of the radial velocity ($|v_r|$), (c) radial normal stress ($\sigma_{rr}$) and (d) tangential normal stress ($\sigma_{\theta\theta}$) with radial distance from the exit ($r-r_0$), along the centerline of the hopper ($\theta_r=0$) for different values of the wall friction ($\mu_w$). The dashed lines are predictions of the theory.}\label{fig:muwr}
\end{center}
\end{figure}

We next consider the effect of \textit{wall friction coefficient}, $\mu_w$. Fig.~\ref{fig:muwr} shows the variation with radial distance from the exit ($r-r_0$), along the centerline of the hopper, of the solid fraction ($\phi$), magnitude of the radial velocity ($|v_r|$), and the normal stresses ($\sigma_{rr}, \sigma_{\theta\theta}$) for different values of the wall friction coefficient, $\mu_w$. In this case, the solid fraction is nearly independent of the friction coefficient (Fig.~\ref{fig:muwr}(a)). The velocity decreases with increasing wall friction coefficient, $\mu_w$ (Fig.~\ref{fig:muwr}(c)), and saturates for $\mu_w\ge0.2$. The velocity at the exit ($r=r_0$) is $|v_r| \approx 6$ for $\mu_w = 0$ and reduces to $|v_r| \approx 4$ for $\mu_w \ge 0.2$. The stresses are significantly higher for $\mu_w = 0.0$ and and decrease with wall friction, and also saturate for $\mu_w \ge 0.2$. The lower values of the stress at higher wall friction are due to the \citet{janssen1895versuche} effect, in which the weight of the particles is partially transferred to the walls by wall friction.

\begin{figure}
\begin{center}
\includegraphics[width=\fw in]{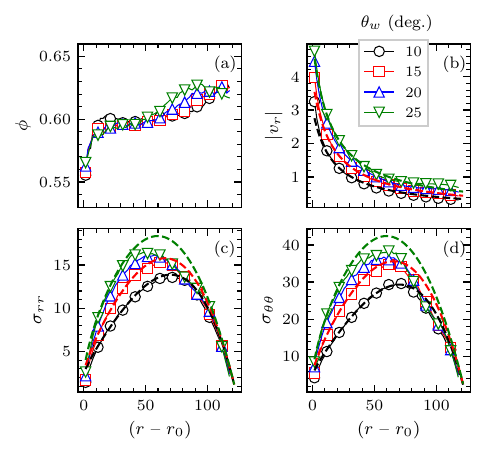}
\caption{Variation of (a) the solid volume fraction ($\phi$), (b) magnitude of the radial velocity ($|v_r|$), (c) radial normal stress ($\sigma_{rr}$) and (d) tangential normal stress ($\sigma_{\theta\theta}$) with radial distance from the exit ($r-r_0$), along the centerline of the hopper ($\theta_r=0$) for different hopper angles ($\theta_w$). The dashed lines are predictions of the theory.}\label{fig:thr}
\end{center}
\end{figure}

The effect of the \textit{wedge angle}, $\theta_w$, (Fig.~\ref{fig:sch}) on the centerline profiles of the solid fraction ($\phi$), radial velocity magnitude ($|v_r|$) and normal stresses ($\sigma_{rr}, \sigma_{\theta\theta}$) is shown in Fig.~\ref{fig:thr}. The radial distance to the exit plane ($r_0$) is kept constant at the base case value, and hence the orifice width ($D$) increases with wedge angle, since $D=2r_0\sin\theta_w$. We kept $r_0$ constant instead of $D$, since the velocity and and stress distributions are independent of the wedge angle for frictionless side walls ($\mu_w = 0$), as shown by \citet{momin2025granular}. In the present case, where the wall friction is $\mu_w = 0.5$, the solid fraction ($\phi$) is nearly independent of wedge angle, however, the velocity magnitude ($|v_r|$) increases. This is due to a lower effective wall frictional force per unit volume of the material, at the larger wedge angles, $\theta_w$. Both normal stresses ($\sigma_{rr}, \sigma_{\theta\theta}$) also increase and the maxima in the stresses shift to lower radii with wedge angle ($\theta_w$), for the same reason. Thus, increasing the wedge angle is equivalent to reducing the wall friction, which shows a similar trend for the velocity and stresses (Fig.~\ref{fig:muwr}), as the present case. This is in contrast to the effect of increasing the orifice width, $D$, in which the velocity increases but the normal stresses are nearly constant (Fig.~\ref{fig:Dr}).

\begin{figure}
\begin{center}
\includegraphics[width=\fw in]{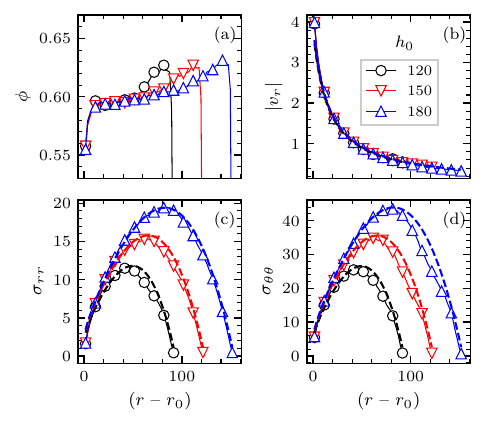}
\caption{Variation of (a) the solid volume fraction ($\phi$), (b) magnitude of the radial velocity ($|v_r|$), (c) radial normal stress ($\sigma_{rr}$) and (d) tangential normal stress ($\sigma_{\theta\theta}$) with radial distance from the exit ($r-r_0$), along the centerline of the hopper ($\theta_r=0$) for different hopper heights ($h_0$). The dashed lines are predictions of the theory.}\label{fig:hr}
\end{center}
\end{figure}

Finally, we consider the effect of \textit{hopper height}, $h_0$, on the centerline profiles, shown in Fig.~\ref{fig:hr}. The solid fraction profiles ($\phi(r)$) are nearly the same for the three heights, except near the free surface, where the solid volume fraction increases slightly, in each case. The velocity profiles ($|v_r(r)|$) for the three heights are nearly identical. The maxima in the normal stress profiles, for both the radial ($\sigma_{rr}(r)$) and tangential ($\sigma_{\theta\theta}(r)$) stresses, increase with height and shift to higher radial positions, but the profiles are similar. In particular, the stress profiles for the different heights are identical near the exit ($r = r_0$). The data for the velocity and stresses are consistent with the well-established results that the mass flow rate and the stress at the exit are independent of height of material in the hopper \cite{tighe2007pressure, janssen1895versuche, beverloo1961flow}.

\begin{figure}
\begin{center}
\includegraphics[width=\fw in]{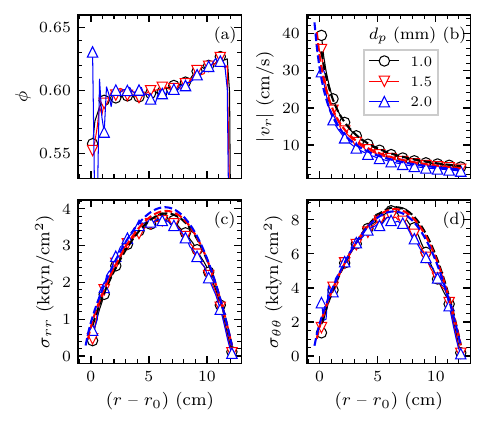}
\caption{Variation of (a) the solid volume fraction ($\phi$), (b) magnitude of the radial velocity ($|v_r|$), (c) radial normal stress ($\sigma_{rr}$) and (d) tangential normal stress ($\sigma_{\theta\theta}$) with radial distance from the exit ($r-r_0$), along the centerline of the hopper ($\theta_r=0$) for different particle diameters ($d_p$). The variables are in dimensional form. The dashed lines are predictions of the theory.}\label{fig:dpr}
\end{center}
\end{figure}

We consider next, the effects of the variation of the particle properties, particle diameter ($d_p$) and particle friction coefficient ($\mu_p$), on the profiles.

Fig.~\ref{fig:dpr} shows the variation of the solid fraction ($\phi$), radial velocity magnitude ($|v_r|$), and normal stresses ($\sigma_{rr}, \sigma_{\theta\theta}$) with distance from the exit ($r-r_0$), along the centerline of the hopper ($\theta_r=0$), for three different \textit{particle diameters}, $d_p$. The variables are given in dimensional form in this set of graphs, with the units indicated.  The results indicate that the solid fraction ($\phi$), and normal stresses ($\sigma_{rr}$, $\sigma_{\theta\theta}$) are nearly independent of the particle diameter, $d_p$. The velocity, however, decreases significantly with particle diameter, with the exit velocity reducing from $|v_r| \approx 40\text{ cm/s}$ to $|v_r| \approx 30\text{ cm/s}$ when the particle diameter is increased from $d_p = 1\text{ mm}$ to $d_p = 2\text{ mm}$. This is consistent with the \citet{beverloo1961flow} correlation (Eq.~(\ref{eq:hag})), as shown below.

\begin{figure}
\begin{center}
\includegraphics[width=\fw in]{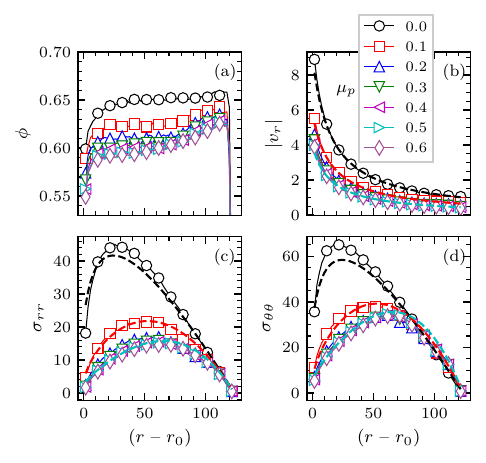}
\caption{Variation of (a) the solid volume fraction ($\phi$), (b) magnitude of the radial velocity ($|v_r|$), (c) radial normal stress ($\sigma_{rr}$) and (d) tangential normal stress ($\sigma_{\theta\theta}$) with radial distance from the exit ($r-r_0$), along the centerline of the hopper ($\theta_r=0$) for different particle friction coefficients ($\mu_p$). The dashed lines are predictions of the theory.}\label{fig:mupr}
\end{center}
\end{figure}

The centerline profiles for $\phi$, $|v_r|$, $\sigma_{rr}$ and $\sigma_{\theta\theta}$ are shown in Fig.~\ref{fig:mupr} for different values of the \textit{interparticle friction coefficient}, $\mu_p$. The solid fraction ($\phi$) is high ($\phi > 0.65$) for frictionless particles ($\mu_p=0$) and reduces with increasing $\mu_p$. The solid fraction profile saturates for $\mu_p \ge 0.4$. The velocity also decreases substantially with increase in particle friction coefficient, with the exit velocity reducing from $|v_r| \approx 9$ to $|v_r| = 4$, when the friction coefficient is increased from $\mu_p = 0.0$ to $\mu_p = 0.2$. The velocity profile saturates for $\mu_p \ge 0.2$. The stresses are higher for low values of the friction coefficient ($\mu_p$) and the maxima are at lower radial positions, but saturate for $\mu_p \ge 0.2$. The reasons for the higher stresses at low $\mu_p$ are again due to a lower transfer of the particle weight to the side walls by friction, according to the \citet{janssen1895versuche} mechanism.

\begin{figure}
\begin{center}
\includegraphics[width=\fw in]{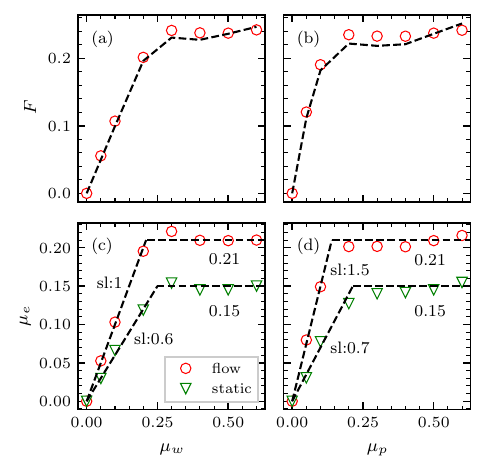}
\caption{Variation of (a), (b) the velocity slope parameter, $F$, and (c), (d) the effective wall friction coefficient, $\mu_e$ with the wall friction coefficient, $\mu_w$ and the wall friction coefficient, $\mu_p$. The dashed lines in (a), (b) are predictions of Eq.~\ref{eq:F} and in (c), (d) are bilinear fits. \label{fig:Fmue}}
\end{center}
\end{figure}

\begin{figure}
\begin{center}
\includegraphics[width=\fw in]{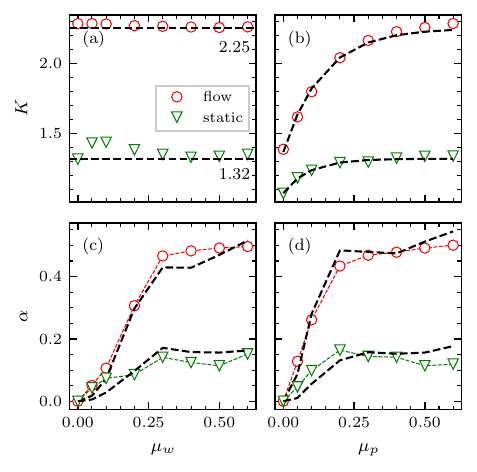}
\caption{Variation of the rheological parameters (a), (b)  $K$, and (c), (d) $\alpha$, with the wall friction coefficient, $\mu_w$ and the wall friction coefficient, $\mu_p$. The dashed lines in (b) are fits of Eq.~(\ref{eq:Kmu}) and in (c), (d) predictions of Eq.~(\ref{eq:almu}).\label{fig:Kal}}
\end{center}
\end{figure}

The scaling relations discussed in Sec.~\ref{sec:sca} are found to be valid for all the cases studied and fitted values of the parameters, $a$, $F$, $\mu_e$, $K$ and $\alpha$, are given in Table~\ref{tab:sys} for the varying system parameters ($D, \mu_w, \theta_w \text{ and } h_0$), and in Table~\ref {tab:part} for the different particle properties ($d_p$, $\mu_p$). The data given in Tables~\ref{tab:sys} and \ref{tab:part} indicate that the velocity parameter, $a$, increases with orifice width, $D$, hopper angle, $\theta_w$, and decreases with wall friction coefficient, $\mu_w$, particle diameter, $d_p$, and particle friction coefficient, $\mu_p$. The parameter, $a$, is nearly independent of hoppers height, $h_0$. The velocity profile slope parameter, $F$, is nearly independent of the orifice width, $D$, hopper height, $h_0$, and particle diameter, $d_p$. The parameter increases with wall friction, $\mu_w$, wedge angle, $\theta_w$, and particle friction, $\mu_p$. The variation of the effective wall friction, $\mu_e$, with the parameters is similar to the velocity profile slope parameter, $F$. The parameter $K$, which is related to the internal friction coefficient, $f$, is nearly constant with $D$, $\mu_w$, $\theta_w$, $h_0$ and $d_p$, and increases with the particle friction, $\mu_p$. Finally, the rheological parameter, $\alpha$, which relates to the decrease in the internal friction, $f$ with angle, $\theta_r$, is nearly constant with $D$, $\theta_w$, $h_0$, and $d_p$, and decreases with both $\mu_w$ and $\mu_p$. Fitting of the scaling relations for the stress ratios was also done for all the static cases, and fitted values of the parameters ($\mu_e$, $K$, and $\alpha$) are given in the supplemental material.

We consider the variation of the parameters $f$, $\mu_e$, $K$ and $\alpha$ with the friction coefficients, $\mu_w$ and $\mu_p$, in more detail, next. The coaxiality condition relates the parameters $F$ and $\mu_e$ as
\begin{equation}
\frac{F}{\theta_w(1-\theta_r^2)}=\frac{\mu_e\kappa}{\kappa-1},
\end{equation}
using the scaling relations (Eqs.~(\ref{eq:scrt}) and (\ref{eq:sctt})). For small $\theta_r$, we have
\begin{equation}\label{eq:F}
F=\frac{2\theta_w\mu_e}{B_F(K-1)},
\end{equation}
where $B_F=(1-\mu_w/3)$ is an empirical factor introduced to account for the deviation of the data from the coaxiality condition. Fig.~\ref{fig:Fmue} shows the variation of $F$ and $\mu_e$ with the friction coefficients, $\mu_w$ and $\mu_p$. The dashed lines in Figs.~\ref{fig:Fmue}(a) and (b) are the predictions of Eq.~(\ref{eq:F}), which match the computational data. Thus, a modified version of the coaxiality condition predicts the variation of the velocity slope parameter, $F$. The dashed lines in Figs.~\ref{fig:Fmue}(c) and (d) are bilinear fits to the data, and show that $\mu_e$ increases linearly with the friction coefficients ($\mu_w$, $\mu_p$), and then saturates to a constant value for both the flowing and static systems. The slopes of the lines are indicated.

Fig.~\ref{fig:Kal} shows the variation of the rheological parameters, $K$ and $\alpha$, with the the friction coefficients ($\mu_w$, $\mu_p$) for the flowing and static systems. The stress ratio, $K$, is nearly constant for varying wall friction coefficient ($\mu_w$) for both static and flowing systems, which is reasonable. $K$, is significantly higher for the flowing system, indicating a higher internal friction coefficient, $f$, as found by \citet{jenike1964}. The stress ratio ($K$) increases exponentially with the particle friction coefficient ($\mu_p$) for static and flowing systems, and saturates to the corresponding value shown in Fig.~\ref{fig:Kal}(c) . The dashed lines are fits of
\begin{equation}\label{eq:Kmu}
K =  K_s-\Delta K\exp(-\mu_p/\mu*),
\end{equation}
where $\mu*$ and $\Delta K$ are fitted constants and $K_s$ are the saturated values at high friction given in Fig~\ref{fig:Kal}(a). The parameter values obtained from fitting are $\Delta K=0.88$ and $\mu*=0.14$ for the flowing system and $\Delta K=0.24$ and $\mu*=0.09$ for the static system.

The variation of $\alpha$ with the particle friction coefficient ($\mu_p$) is shown in Figs.~\ref{fig:Kal}(c), (d). $\alpha$ increases with both $\mu_w$ and $\mu_p$ for the flowing and static systems, and then saturates to a constant value. The dashed lines are predictions of a modified version of Eq.~(\ref{eq:al}) given by
\begin{equation}\label{eq:almu}
\alpha=B_{\alpha}\mu_e^2K(K+1)/(K-1),
\end{equation}
where we introduce an empirical factor $B_{\alpha}$, with $B_{\alpha}=(1+\mu_w/0.6)$ for the flowing system and $B_{\alpha}=0.9$ for the static system, for variation with $\mu_w$ and $\mu_p$.

\begin{figure}
\begin{center}
\includegraphics[width=\fw in]{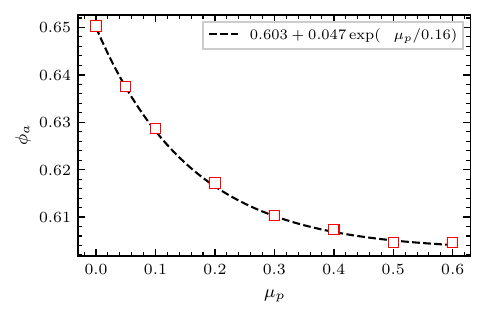}
\caption{Variation of the average solid volume fraction ($\phi_a$) with interparticle friction coefficient, $\mu_p$. The dashed line is a fit of the function indicated in the legend. The average is over radial positions $(r-r_0)\in(20,115)$. \label{fig:php}}
\end{center}
\end{figure}

 We use the fitted values of the constants of the scaling relations: $a$, $F$, $\mu_e$, $K$ and $\alpha$, given in Tables~\ref{tab:sys} and \ref{tab:part} to obtain predictions of the theory for all the cases studied, which are shown as dashed lines in Figs.~\ref{fig:Dr}-\ref{fig:mupr}. The solid fraction used in all the cases was $\phi=0.6$, except for the case of varying interparticle friction, where the corresponding higher values were used, and are shown in Fig.~\ref{fig:php}. The predictions of the theory match the simulation results in all cases, except the following. The theory overpredicts the stresses at the highest wedge angle ($\theta_w=25^{\circ}$, Fig.~\ref{fig:thr}(c), (d)) and underpredicts the stresses at the lowest interparticle friction ($\mu_p=0$, Fig.~\ref{fig:mupr}(c), (d)). Deviations between theory and simulation results also occur for the tangential normal stress in the upper region of the hopper for the largest hopper height ($h_0=180$, Fig.~\ref{fig:hr}(d)).

\begin{table*}
\caption{Fitted parameter values for the scaling relations for velocity ($a,F$), stress ratios ($\mu_e,K,\alpha$) and the time variation of the mass flow rate ($\dot{m}_s,\tau$) for varying system parameters ($D,\mu_w,\theta_w,h_0$). The mass flow rates obtained from simulations ($\dot{m}$) are also shown.}\label{tab:sys}
\begin{center}
\input tab1.tex
\end{center}
\end{table*}

\begin{table*}
\caption{Fitted parameter values for the scaling relations for velocity ($a,F$), stress ratios ($\mu_e,K,\alpha$) and the time variation of the mass flow rate ($\dot{m}_s,\tau$) for varying particle properties ($d_p,\mu_p$). The mass flow rates obtained from simulations ($\dot{m}$) are also shown.}\label{tab:part}
\begin{center}
\input tab2.tex
\end{center}
\end{table*}

\subsection{Mass flow rate}\label{sec:md}
The variation of the mass flow rate with flow time, similar to Fig.~\ref{fig:mdt}, was plotted for all the cases, and the variation was exponential in all cases. Eq.~\ref{eq:mdt} was fitted to each data set and the fitted values ($\dot{m}_s, \tau$) are given in Tables~\ref{tab:sys} and \ref{tab:part}. The steady state mass flow rate ($\dot{m}$) for each case is also shown and we have $\dot{m} \approx \dot{m}_s$, for all the cases. The variation of the mass flow rate with the parameters is the same as that for the velocity parameter, $a$, discussed above. The characteristic time for the flow ($\tau$) increases with orifice width ($D$), but $\tau$ is nearly constant ($\tau \approx 2$) with $\mu_w, \theta_w, h_0$. The characteristic time reduces with particle diameter ($d_p$) and is constant with particle friction ($\mu_p$).

\begin{figure}
\begin{center}
\includegraphics[width=\fw in]{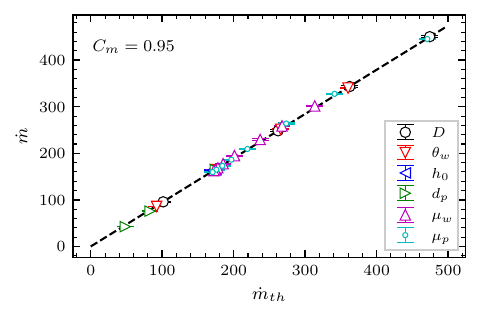}
\caption{Comparison of the mass flow rate ($\dot{m}$) obtained from simulations to the predictions of the mass flow rate from Eq.~(\ref{eq:mth}) (($\dot{m}_{th}$). \label{fig:mdall}}
\end{center}
\end{figure}

We consider the theoretical predictions of the mass flow rate, next. The mass flow rate based on the velocity correlation is given by
\begin{equation}\label{eq:mth}
\dot{m}_{th}=2\phi b\int_0^{\theta_w}rv_r d\theta=2\phi ab\theta_w(1-F/3),
\end{equation}
where $r$ is any radial position. Fig.~\ref{fig:mdall} shows the variation of the mass flow rate from simulations, $\dot{m}$, to the predictions of Eq.~(\ref{eq:mth}), $\dot{m}_{th}$, using the fitted values of $a$ and $F$. The data fall on a straight line, and the dashed line is a fit of $\dot{m}=C_m\dot{m}_{th}$ to the data, with $C_m=0.94$. The small deviation of $C_m$ from unity is because of the deviation of the velocity from the scaling relation near the wall ($\theta_r=1$).

We obtain an estimate for the velocity parameter, $a$, using a stress boundary condition at the exit of the hopper. The radial normal stress along the centerline ($\theta=0$) at the exit ($r=r_0$), is obtained from Eq.~(\ref{eq:srra}) as
\begin{equation}
\begin{split}
\sigma_{rr}(r_0, 0) &= \frac{\phi r_0}{p-1} \left[ 1 - \left(\frac{r_0}{R}\right)^{p-1} \right]\\& - \frac{\phi a^2}{(p+2)r_0^2} \left[ 1 - \left(\frac{r_0}{R}\right)^{p+2} \right]
\end{split}
\end{equation}
with $p = K(1+\mu_e/\theta_w)-1$, which reduces to
\begin{equation}
\sigma_{rr}(r_0, 0) = \frac{\phi r_0}{p-1} - \frac{\phi a^2}{(p+2)r_0^2},
\end{equation}
for $p > 2$ and $r_0/R \ll 1$. In all the cases of high friction ($\mu _{p} > 0 . 2$), we have $p \approx 4$ and $r / R _{0} < 0.1$, so the approximation is justified. For $\mu_p \leq 0.05$ the factor $[ 1 - ( r/R _{0}) ^{p - 1}]$ is less than unity and is included in the calculation of the stress at the exit. Assuming that the exit stress is proportional to the first term, $\sigma_{rr} \propto \phi r_0/(p-1)$, we obtain
\begin{equation}\label{eq:a}
a = C \left( \frac{p+2}{p-1} \right)^{1/2} r_0^{3/2},
\end{equation}
where $C < 1$ takes into account the non-zero exit stress. 

The estimate of the velocity parameter ($a$) enables the calculation of the mass flow rate in terms of the system parameters. Substituting for $a$ in Eq.~(\ref{eq:mth}), and using $r_0 = D/(2\sin \theta_w)$, we obtain
\begin{equation}
\dot{m}_{th} = \frac{C \phi b \theta_w (1-F/3)}{\sqrt{2}(\sin\theta_w)^{3/2}} \left( \frac{p+2}{p-1} \right)^{1/2} D^{3/2}. \end{equation}
The velocity at the hopper walls is lower than that predicted by the velocity correlation, as shown in Fig.~\ref{fig:phvr}. Incorporating this lower velocity at the walls of the hopper in terms of a reduced orifice width, $D$, we obtain the mass flow rate in the form of the Beverloo correlation \cite{beverloo1961flow} as
\begin{equation}\label{eq:bev}
\dot{m}_{th} = C G (D-k)^{3/2},
\end{equation}
where
\begin{equation}
G = \frac{\phi b \theta_w (1-F/3)}{\sqrt{2} (\sin \theta_w)^{3/2}} \left( \frac{p+2}{p-1} \right)^{1/2},
\end{equation}
and $k$ is the dimensionless reduction in the orifice width.

\begin{figure}
\begin{center}
\includegraphics[width=\fw in]{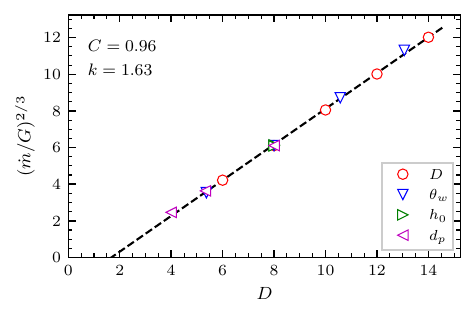}
\caption{Beverloo \cite{beverloo1961flow} plot showing the variation of $(\dot{m}/G)$ with orifice width, $D$ for different parameters as indicated in the legend The dashed line is a fit of Eq.~(\ref{eq:bev}) to the data and the fitted values of the constants are given in the figure. \label{fig:mdD}}
\end{center}
\end{figure}

Fig.~\ref{fig:mdD} shows the variation of the mass flow rate scaled by the factor $G$ and are computed values of $(\dot{m}/G)^{2/3}$ with orifice width, $D$, for different values of angle, $\theta_w$, height, $h_0$, and particle diameter, $d_p$. All the data collapse to a single straight line and the dashed line is a fit to the data, with $C=0.96$ and $k=1.63$. The value of $k$ is close to the reported values. The collapse of the data to a straight line implies that assumption of the exit stress being proportional to $\phi r_0 / (p-1)$ is reasonable for varying $D$, $\theta_w$, $h_0$ and $d_p$.

\begin{figure}
\begin{center}
\includegraphics[width=\fw in]{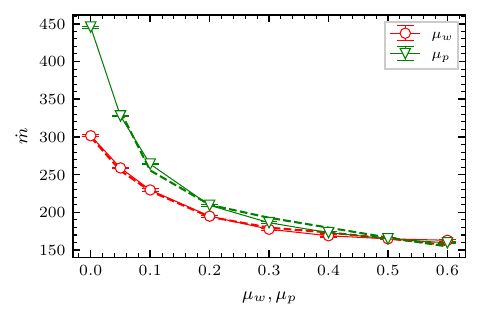}
\caption{Variation of the mass flow rate, $\dot{m}$, with the wall friction coefficient, $\mu_w$ and the particle friction coefficient, $\mu_p$. The dashed lines are predictions of Eq.~(\ref{eq:bev}) with the constant $C$ given by Eq.~(\ref{eq:C}). \label{fig:mdmu}}
\end{center}
\end{figure}

For varying friction coefficients, the constant $C$ varies significantly with the wall friction coefficient ($\mu_w$) and the particle friction coefficient ($\mu_p$), and is given by
\begin{equation}\label{eq:C}
C = 1.5 - 0.48\mu_w - 0.59\mu_p,
\end{equation}
for $\mu_w,\mu_p\le0.6$. Fig.~\ref{fig:mdmu} shows a comparison of the predicted mass flow rate ($\dot{m}_{th}$, Eq.~(\ref{eq:bev})) with the computed mass flow rate ($\dot{m}$, symbols) for varying friction coefficients ($\mu_w, \mu_p$), using the constant $C$ given in Eq.~(\ref{eq:C}). There is very good agreement between the two. The theory significantly overpredicts the mass flow rate for frictionless particles ($\mu_p=0$), and is not shown. The dependence of $C$ on the friction coefficient is due to the variation of the exit stress with friction coefficients, different from that in the term $\phi r_0/(p-1)$.

We note that the factor, $G$, in the above analysis is computed using parameters evaluated in the bulk of the flow. The parameters have different values at the exit. Here we assume that the differences may be incorporated in the constant, $C$.

\subsection{Larger than gravity acceleration}\label{sec:acc}
\begin{figure}
\begin{center}
\includegraphics[width=\fw in]{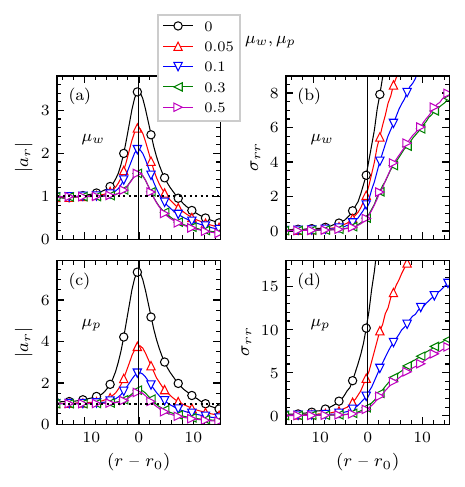}
\caption{Variation of (a), (c) the magnitude of the acceleration ($|a_r|$), and (b), (d) the normal stress ($\sigma_{rr}$) with height ($r-r_0$) for different values of the friction coefficients ($\mu_w, \mu_p$). \label{fig:muva}}
\end{center}
\end{figure}

In all the cases studied, the acceleration of the particles in the exit region is significantly larger than the acceleration due to gravity. The maximum acceleration of the particles does not vary significantly with the parameters $D$, $\theta_w$, $h_0$ and $d_p$. However, the variation of the maximum acceleration with friction coefficients  is quite large, as shown in Figs.~\ref{fig:muva} (a), (c). The maximum acceleration increases significantly when friction coefficients are reduced, and the maximum acceleration is $|a_y| \approx 3.5$ for frictionless walls ($\mu_w=0$) and is $|a_y| \approx 7$ for frictionless particles ($\mu_p=0$). The higher than gravity acceleration is due to the high normal gradients ($\partial\sigma_{yy}/\partial y$) near the exit (Figs.~\ref{fig:muva} (b), (d)). Such gradients do not occur in fluids flowing out of orifices, since the fluid flows into a volume at atmospheric pressure. In contrast, granular flow from orifices occurs into a vacuum (zero granular pressure), which facilitates the flow out of the hopper. Lower wall friction and particle friction values enhance this effect, resulting in higher accelerations. 

The effect described above is incorporated in the theory via the boundary condition, $\sigma_{rr}\approx0$ at $r=r_0$, that is a near vacuum at the exit, and the theory predicts the high normal stress gradient at the exit. The scaling of the mass flow rate remains the same, however, the lower friction values increase the magnitude of the constant $C$.

\section{Conclusions}\label{sec:con}
Results of a computational study of granular flows in a wedge-shaped hopper,  based on DEM simulations, are presented for varying system parameters ($D$,  $\theta_w$, $h_0$, $\mu_w$) and particle properties ($d_p$, $\mu_p$). The data for the solid volume fraction, velocity and stress distributions provide a  comprehensive view of the behaviour in the hopper, for the flowing and static systems.

Scaling relations are proposed for the radial velocity, based on a perturbation solution and prior results, and for the stress ratios based on expressions for the stress components in terms of \citet{sokolovski1960statics} variables.  The velocity is found to be purely radial, with the velocity given by
\begin{equation}
v_r = \frac{a}{r}[1 - F\theta^2],
\end{equation}
where $a$ and $F$ are constants. The stress components are related to each other by the following scaling relations for the stress ratios
\begin{eqnarray}
{\sigma_{r\theta}}/{\sigma_{\theta\theta}} &=&-\mu_e \theta_r, \\
{\sigma_{\theta\theta}}/{\sigma_{rr}} &=& K(1 - \alpha \theta_r^2),
\end{eqnarray}
where $\mu_e$, $K$ and $\alpha$ are constants. The scaling relations are found to be valid for all the cases, and values of the fitted values are reported. 

The primary model parameters are, $K$, related to the internal friction  coefficient, and, $\mu_e$, the effective wall friction. Both parameters are constant in the bulk of the hopper. Expressions to compute the other parameters ($a, F, \alpha$) are given. The parameters vary primarily with wall friction and particle friction coefficients ($\mu_w, \mu_p$) and saturate at moderate values of the friction coefficients ($\mu_w, \mu_p > 0.3$) to constant values. For the flowing system the parameters are: $F = 0.24$,  $\mu_e = 0.21$, $K = 2.25$ and $\alpha = 0.49$, and for the static system, the parameters are: $\mu_e = 0.15$, $K = 1.32$ and $\alpha = 0.18$. The velocity parameter, $a$, is given by Eq.~(\ref{eq:a}). 

The internal friction coefficient, $f$, computed from data, is independent of the inertial number (dimensionless shear rate) indicating the validity of the Mohr--Coulomb rheology. However, the internal friction coefficient varies spatially. Further, the coaxiality condition is approximately satisfied (within $2^\circ$) over most of the domain but deviates near the hopper wall ($\theta_r > 0.9$).  The effective wall friction, $\mu_e$, is lower than the wall friction coefficient, $\mu_w$.

A theory is presented to compute the stresses using the scaling relations in place of the constitutive relations. Good agreement is obtained between predictions of the theory and simulation results for all the cases studied. An expression for the mass flow rate, which is a modified form of the Beverloo correlation \cite{beverloo1961flow}, is obtained from the theory.  The expression has a prefactor ($G$) dependent on the parameters, and a constant $C =0.96$, for the moderate friction cases. For the low friction cases, the same relation holds, but with the constant, $C(\mu_w,\mu_p)>1$, and decreasing linearly with increasing friction. The expression gives good predictions for all the different cases studied, including for varying friction coefficients.

Finally, the acceleration of particles is larger than gravity near the exit for all the cases, with the maximum acceleration is as high as $a_y \approx 7g$ for frictionless particles. The cause of the larger than gravity acceleration is the high normal stress gradient, essentially due to the hopper discharging in a granular vacuum.

The parameters of the scaling relations vary with radial distance in the exit region, most likely due to collisional stresses. Developing a rheological model that accounts for the shear rate dependence of stresses in the exit region, as well as, the spatially varying friction coefficient and non-coaxiality in the entire hopper, is the subject of future work.

\bibliography{ref}
\end{document}

%% file: tab1.tex
\begin{tabular}{|c|c|c|c|c|c|c|c|c|}
\hline
$D$ &$a$ &$F$ &$\mu_e$ &$K$ &$\alpha$ &$\dot{m}_s$ &$\tau$ &$\dot{m}$ \\ \hline
6 &35.0 &0.24 &0.21 &2.19 &0.48 &96.3 &1.71 &$95.8\pm0.7$ \\ 
8 &60.3 &0.24 &0.21 &2.26 &0.49 &165.4 &1.97 &$165.0\pm1.3$ \\ 
10 &90.6 &0.24 &0.21 &2.29 &0.50 &249.2 &2.33 &$248.4\pm1.8$ \\ 
12 &125.1 &0.23 &0.21 &2.31 &0.49 &345.3 &2.58 &$343.5\pm2.0$ \\ 
14 &163.9 &0.24 &0.21 &2.31 &0.48 &454.4 &2.87 &$450.5\pm2.6$ \\ 
\hline
$\mu_w$ &$a$ &$F$ &$\mu_e$ &$K$ &$\alpha$ &$\dot{m}_s$ &$\tau$ &$\dot{m}$ \\ \hline
0.00 &99.8 &0.00 &0.00 &2.28 &0.00 &303.7 &2.10 &$301.6\pm1.6$ \\ 
0.05 &86.8 &0.06 &0.05 &2.28 &0.05 &259.2 &2.11 &$258.9\pm1.2$ \\ 
0.10 &78.2 &0.11 &0.10 &2.28 &0.11 &229.8 &2.11 &$229.7\pm1.7$ \\ 
0.20 &68.6 &0.20 &0.20 &2.27 &0.31 &195.2 &2.05 &$195.0\pm1.2$ \\ 
0.30 &64.1 &0.24 &0.22 &2.26 &0.47 &177.9 &2.02 &$177.6\pm1.3$ \\ 
0.40 &61.5 &0.24 &0.21 &2.26 &0.48 &169.1 &1.97 &$169.0\pm1.6$ \\ 
0.50 &60.3 &0.24 &0.21 &2.26 &0.49 &165.4 &1.97 &$165.0\pm1.3$ \\ 
0.60 &59.7 &0.24 &0.21 &2.26 &0.50 &163.2 &2.00 &$163.0\pm1.2$ \\ 
\hline
$\theta_w$ (deg.) &$a$ &$F$ &$\mu_e$ &$K$ &$\alpha$ &$\dot{m}_s$ &$\tau$ &$\dot{m}$ \\ \hline
10 &46.9 &0.18 &0.21 &2.11 &0.54 &86.3 &1.97 &$85.9\pm0.9$ \\ 
15 &60.3 &0.24 &0.21 &2.26 &0.49 &165.4 &1.97 &$165.0\pm1.3$ \\ 
20 &69.9 &0.28 &0.21 &2.28 &0.44 &250.9 &2.02 &$251.6\pm1.5$ \\ 
25 &77.1 &0.33 &0.21 &2.31 &0.42 &340.7 &2.13 &$340.0\pm1.5$ \\ 
\hline
$h_0$ &$a$ &$F$ &$\mu_e$ &$K$ &$\alpha$ &$\dot{m}_s$ &$\tau$ &$\dot{m}$ \\ \hline
120 &58.9 &0.24 &0.23 &2.25 &0.49 &164.7 &2.03 &$163.7\pm1.4$ \\ 
150 &60.3 &0.24 &0.21 &2.26 &0.49 &165.4 &1.97 &$165.0\pm1.3$ \\ 
180 &60.4 &0.23 &0.20 &2.26 &0.49 &165.8 &1.95 &$165.3\pm1.0$ \\ 
\hline
\end{tabular}

%% file: tab2.tex
\begin{tabular}{|c|c|c|c|c|c|c|c|c|}
\hline
$d_p$ (mm) &$a$ &$F$ &$\mu_e$ &$K$ &$\alpha$ &$\dot{m}_s$ &$\tau$ &$\dot{m}$ \\ \hline
1.0 &60.3 &0.24 &0.21 &2.26 &0.49 &165.4 &1.97 &$165.0\pm1.3$ \\ 
1.5 &28.8 &0.25 &0.21 &2.19 &0.53 &75.8 &1.25 &$75.9\pm0.8$ \\ 
2.0 &16.8 &0.26 &0.22 &2.10 &0.53 &42.4 &0.93 &$42.6\pm0.6$ \\ 
\hline
$\mu_p$ &$a$ &$F$ &$\mu_e$ &$K$ &$\alpha$ &$\dot{m}_s$ &$\tau$ &$\dot{m}$ \\ \hline
0.00 &138.3 &0.00 &0.00 &1.39 &0.00 &453.0 &2.02 &$445.5 \pm1.5$\\ 
0.05 &106.4 &0.12 &0.08 &1.62 &0.13 &330.2 &2.06 &$327.8\pm1.0$ \\ 
0.10 &88.7 &0.19 &0.15 &1.80 &0.26 &264.7 &2.07 &$264.0 \pm1.1$\\ 
0.20 &73.6 &0.23 &0.20 &2.04 &0.43 &209.8 &2.07 &$209.4 \pm1.2$\\ 
0.30 &66.7 &0.23 &0.20 &2.16 &0.47 &186.5 &2.03 &$186.3 \pm1.1$\\ 
0.40 &62.8 &0.23 &0.20 &2.23 &0.48 &173.1 &2.03 &$173.2 \pm1.4$\\ 
0.50 &60.3 &0.24 &0.21 &2.26 &0.49 &165.4 &1.97 &$165.0 \pm1.3$\\ 
0.60 &58.6 &0.24 &0.22 &2.28 &0.50 &160.0 &2.03 &$159.6 \pm1.2$\\ 
\hline
\end{tabular}